\documentclass[]{aa}  

\usepackage{natbib}
\bibpunct{(}{)}{;}{a}{}{,}
\usepackage{graphicx}
\usepackage{revsymb}	
\usepackage{amsmath}	
\usepackage[usenames]{color}	
\usepackage{epstopdf}	
\usepackage{txfonts}
\usepackage{amssymb}
\usepackage{color}
\usepackage[justification=centering]{caption}
\usepackage{geometry} 
\usepackage[parfill]{parskip} 
\usepackage{amssymb}
\usepackage{slantsc}
\usepackage{array}
\usepackage{url}
\usepackage{amsfonts}
\usepackage{hyperref} 
\usepackage{sidecap}
\usepackage{nicefrac}
\usepackage{subfigure}
\usepackage{epsfig}
\usepackage{xcolor}
\colorlet{bleu}{blue!90!darkgray}
\colorlet{rouge}{red!70!darkgray}
\colorlet{vert}{green!60!black}
\colorlet{gris}{black!50}
\begin{document}

\title{Stellar acoustic radii, mean densities and ages from seismic inversion techniques}
\author{G. Buldgen\inst{1}\and D. R. Reese\inst{2}\and M. A. Dupret\inst{1}\and R. Samadi\inst{3}}
\institute{Institut d’Astrophysique et Géophysique de l’Université de Liège, Allée du 6 août 17, 4000 Liège, Belgium \and School of Physics and Astronomy, University of
Birmingham, Edgbaston, Birmingham, B15 2TT, UK. \and LESIA, Observatoire de Paris, CNRS UMR 8109, UPMC, Université Denis Diderot, 5 place Jules Janssen, 92195 Meudon Cedex,
France}
\date{July 16,2014}
\abstract{Determining stellar characteristics such as the radius, the mass or the age is crucial when studying stellar evolution, exoplanetary systems or characterising stellar populations in the Galaxy. Asteroseismology is the golden path to accurately obtain these characteristics. In this context, a key question is how to make these methods less model-dependant.}{Building on the work of \citet{Reese}, we wish to extend the SOLA inversion technique to new stellar global characteristics in addition to the mean density. The goal is to provide a general framework in which to estimate these characteristics as accurately as possible in low mass main sequence stars.}{
First, we describe our framework and discuss the reliability of the inversion technique and the possible sources of error. We then apply this methodology to the acoustic radius, an age indicator based on the sound speed derivative and the mean density and compare it to estimates based on the average large and small frequency separations. These inversions are carried out for several test cases which include: various metallicities, different mixing-lengths, non-adiabatic effects and turbulent pressure.}{We observe that the SOLA method yields accurate results in all test cases whereas results based on the large and small frequency separations are less accurate and more sensitive to surface effects and structural differences in the models. If we include the surface corrections of \citet{Kjeldsen}, we obtain results of comparable accuracy for the mean density. Overall, the mean density and acoustic radius inversions are more robust than the inversions for the age indicator. Moreover, the current approach is limited to relatively young stars with radiative cores. Increasing the number of observed frequencies improves the reliability and accuracy of the method.}{}
\keywords{Stars: interiors -- Stars: oscillations -- Stars: fundamental parameters -- Asteroseismology}
\maketitle
\section{Introduction}


Determining stellar global characteristics such as mass, radius or age as accurately as possible is crucial for understanding stellar evolution, determining properties of exoplanetary systems or characterising stellar populations in the galaxy. Although these quantities can be estimated using classical observations, such as photometry and spectroscopy, or in special cases such as binary systems, significant progress has only been made in recent years with the advent of high precision asteroseismology missions, namely CoRoT and Kepler. Indeed, these missions are providing a wealth of data of unprecedented quality for large numbers of stars. Hence, it is crucial to develop techniques that are able to determine global stellar parameters from pulsation data as accurately as possible and with the least computational effort \citep[see][for a review on this topic]{Chaplin}.
\\
\\
Estimating stellar ages is the most problematic case since there is no direct observational method to measure this quantity. Therefore, it has to be estimated by relating the evolutionary stage empirically to some phenomena like rotation, activity, lithium depletion or by using model-dependent methods like isochrone placements \citep[see][for an extensive review of age determination methods]{Soderblom}. Currently, the most promising method to determine stellar ages is by carrying out asteroseismic modelling of stars. These ages are estimated to be $\sim 10 \%$ accurate in the best cases \citep{Soderblom}.
\\
\\
Many of the techniques used for exploiting stellar pulsation data are variants of grid/parameter search methods. On one end of the spectrum, there are simple methods which estimate global stellar parameters such as the mass and radius through empirical scaling relations based on seismic indicators such as the large frequency separation and the frequency at maximum power. Search methods using a dense grid of models, calculated once and for all, can also be used to find optimal models for a whole set of observed stars. However, it is clear that this method can only handle a limited number of free parameters when describing the models. On the other end of the spectrum, there are sophisticated search methods such as genetic algorithms \citep{Charpinet, Metcalfe} or MCMC methods \citep{Bazot} which are able to deal with much larger multi-dimensional parameter spaces thanks to an optimised search strategy. These methods will typically calculate stellar models as needed, which make them considerably slower than scaling relations or simple grid search methods, thereby limiting the number of observed stars which can be treated this way. A common point in these search methods is their reliance on stellar models which unfortunately do not fully represent the physical complexity of the phenomena taking place in stars. Hence, these inaccuracies can lead to biases in the results and to persistent differences between the model and observed frequencies. Therefore, there is currently a need for methods which are able to characterise the global parameters and evolutionary stage of a star which are less model-dependant, as accurate as possible, and applicable to a large number of stars.
\\
\\
In this context, seismic inversion techniques become particularly interesting since they are able to invert the differences between observed and theoretical frequencies and translate them into appropriate structural corrections on the models. In that sense, these techniques overcome the limitations imposed by the set of physical ingredients used for the construction of the models. Therefore, they allow us to obtain more detailed information on the stellar structure as well as insights into new physical phenomena which need to be included in the models. For instance, helioseismic inversions have provided detailed solar rotation profiles, which were different from theoretical predictions, and have shown that the solution to the lacking solar neutrino problem should come from improving neutrino physics rather than revising the solar structure. In contrast to the solar case, asteroseismic space missions cannot resolve the objects they observe and hence are limited to low degree modes. As a result, it is difficult to obtain reliable inversions of full structural profiles for stars other than the sun. A useful alternative is to invert for global stellar properties. Recently, \citet{Reese} showed how this could be done for the mean density of a star. Such an approach represents an important step of progress compared to using typical seismic indices for two reasons. Firstly, it can provide custom-made global quantities which are directly related to the stellar structure rather than to the pulsation spectra of the stars. Secondly, the associated averaging kernels which are obtained as a by-product give useful indications on how accurate the result is.
\\
\\
In the current paper, we wish to extend this approach to other stellar quantities, namely the acoustic radius and an age indicator based on the integral of the sound speed derivative. These characteristics are not chosen fortuitously. Indeed, they allow us to compare our inversion results with those obtained by current asteroseismic proxies, the large frequency separation and the small frequency separation \citep{Vandakurov, Tassoul}. The outline of the paper will be the following: we will define our general approach to the specific inverse problem of global characteristics in Sect. \ref{generalapproach}. Section \ref{acousticage} will show how this methodology applies to the acoustic radius and the age indicator. Sections \ref{gridtest} and \ref{forwardmodel} will present inversion results for different tests cases. In Sect. \ref{gridtest}, we use the model grid of \citet{Reese}, chosen without any optimisation process\footnote{\citep[see][Section $6$ for further details on this particular point]{Reese}}, to carry out a first series of tests and conclude that an optimization process is necessary to choose the appropriate reference model for each inversion. We present such a method in Sect. \ref{forwardmodel} and test it in different cases which include: changes to the metallicity, modifications to the mixing length parameter, non-adiabatic effects in the frequencies and the effects of turbulent pressure. These test cases are chosen to illustrate current limitations and uncertainties in stellar modelling i.e. the uncertainties in the convection treatment, here mimicked by a mixing-length coefficient mismatch, the uncertainties in chemical composition, mimicked by a metallicity changes, the intrinsic non-adiabaticity of stellar oscillations and the unknown surface effects such as turbulent pressure. Each test case is carried out separately to isolate any effects that the inversion could not correct. We show that using inversion techniques on a appropriate reference model can improve the accuracy with which global stellar characteristics are determined in that it provides accurate results in all these cases. Section \ref{conclusion} summarises our results and discusses the strengths and weaknesses of the method. 

\section{General Approach}\label{generalapproach}
\subsection{Inverse problems and ways of solving them}
As stated in the introduction, we seek to establish a new framework for linear inversion techniques that allows us to determine stellar global characteristics. As for any inversion carried out, our method needs a reference model, an observed star and their respective oscillation frequencies. The reference model has to be close enough to the observational target so that the relation between their relative frequency differences and their structure differences can be deduced from the variational principle. This leads to the following typical linear form:
\begin{align}
\frac{\delta \nu_{n,\ell}}{\nu_{n,\ell}}=&\int_{0}^{1} K^{n,\ell}_{s_{1},s_{2}}\frac{\delta s_{1}}{s_{1}}dx + \int_{0}^{1} K^{n,\ell}_{s_{2},s_{1}}\frac{\delta s_{2}}{s_{2}}dx +\frac{\mathcal{G}(\nu)}{Q_{n,\ell}}, \label{eqfreqstruc}
\end{align}
where $s_{1}$ and $s_{2}$ are structural variables like $\rho_{0}$, $\Gamma_{1}$, $c^{2}$, $u_{0}=P_{0}/ \rho_{0}$, etc... As we will see in the next section, choosing the right couple of variables for the right inversion is not always straightforward. The function $\mathcal{G}(\nu)$ is an ad-hoc correction for the surface term assumed to be a slowly varying function depending only on the frequency. It is usually expressed as a sum of Legendre polynomials and normalised by the factor $Q_{n,l}$, which is the mode inertia normalised by the inertia of a radial mode interpolated to the same frequency \citep{JCD86}. The functions $K^{n,l}_{s_{i},s_{j}}$ are the inversion kernels, derived from the reference model and its eigenmodes \citep{Gough}. One should notice that the behaviour of the kernels is critical to ensure a successful linear inversion, especially when working with asteroseismic targets where the number of frequencies is rather small compared to helioseismic inversions.
\\
\\
The symbol $\delta s/s$ denotes the relative difference between the value of $s$ for the reference model and the target at a given $x=\frac{r}{R}$. In the present work, we use the classical definition of the relative differences between target and model:
\begin{align}
\frac{\delta s}{s}=\frac{s_{\mathrm{obs}}-s_{\mathrm{ref}}}{s_{\mathrm{ref}}}. \label{eqtargetrefstruct}
\end{align}
Other definitions were sometimes used in the past for helioseismic inversions \cite[see][]{Antia94} but were not used in this study. 
\\
\\
It is well known that the inversion problem is ill-posed and that the quality of the inversion (in terms of accuracy but also of reliability) depends critically on the quantity and the accuracy of available data. Therefore in the asteroseismic context, inversions of structural profiles such as the density, the sound speed or even the helium abundances are out of reach for linear inversion techniques. However, we can still make a compromise and search for global quantities.
\\
\\
The SOLA inversion method \citep{Pijpers} naturally lends itself to obtaining global quantities. When using SOLA, we build a linear combination of the inversion kernels that matches a pre-defined target. In other words, we wish to determine the values of the coefficients of the linear combination of frequency differences that will give us information about one global characteristic of the observed target. Using Eq. (\ref{eqfreqstruc}), we can define a target $\mathcal{T}$, which can be any function of $x=\frac{r}{R}$. For example, let us assume we wish to determine the value of a global characteristic $A_{obs}$ the relative perturbation of which is defined by:
\begin{align}
\frac{\delta A_{\mathrm{obs}}}{A}=&\int_{0}^{1}\mathcal{T}(x)\frac{\delta s_{1}}{s_{1}}dx+\int_{0}^{1}\mathcal{T}_{\mathrm{cross}}(x)\frac{\delta s_{2}}{s_{2}}dx.
\end{align}
Assuming that Eq. (\ref{eqfreqstruc}) is satisfied for our model and our target, we wish to build the linear combination of frequency differences such that:
\begin{align}
\sum_{i}c_{i}\frac{\delta \nu_{i}}{\nu_{i}}&=\int_{0}^{1}\mathcal{T}(x)\frac{\delta s_{1}}{s_{1}}dx+\int_{0}^{1}\mathcal{T}_{\mathrm{cross}}(x)\frac{\delta s_{2}}{s_{2}}dx \nonumber \\
&= \frac{\delta A_{\mathrm{obs}}}{A}. \label{eqrealvalue}
\end{align}
This is of course an ideal scenario. For real inversions, the result is more likely to be an estimate $\delta A_{\mathrm{inv}}/ A$ which is expressed as follows:
\begin{align}
\frac{\delta A_{\mathrm{inv}}}{A}=&\int_{0}^{1}K_{avg}(x)\frac{\delta s_{1}}{s_{1}}dx +\int_{0}^{1}K_{\mathrm{cross}}(x)\frac{\delta s_{2}}{s_{2}}dx \nonumber \\&+ \sum_{i}c_{i}\frac{\mathcal{G}(\nu_{i})}{Q_{i}}. \label{eqinvresult}
\end{align}
The functions $K_{\mathrm{avg}}(x)$ and $K_{\mathrm{cross}}(x)$ are the so-called averaging and cross-term kernels and the third term accounts for surface effects. The averaging and cross-term kernels are directly related to the structural kernels of Eq. (\ref{eqfreqstruc}) by the inversion coefficients:
\begin{align}
K_{\mathrm{avg}}(x) &= \sum_{i}c_{i}K^{i}_{s_{1},s_{2}}  (x), \label{eqaveraging}\\
K_{\mathrm{cross}}(x) &= \sum_{i}c_{i}K^{i}_{s_{2},s_{1}}(x) \label{eqcrossterm}.
\end{align}
Thus, in order for the inversion to be accurate, these kernels need to be as close as possible to their respective target functions. One should note that the cross-term kernel will always be present in an inversion result, as a direct consequence of Eq. (\ref{eqfreqstruc}). If $A$ is only related to $s_{1}$, the function $\mathcal{T}_{\mathrm{cross}}$ is simply $0$. In this particular case, the contribution of the integral of $s_{2}$ in Eq. (\ref{eqfreqstruc}) has to be eliminated. When using the SOLA method, we build a cost function \citep[see][for the original definition of the OLA cost function and its analysis in the context of Geophysics]{Backus}: 
\begin{align}
\mathcal{J}_{\mathrm{A}} =& \int_{0}^{1}\left[ K_{\mathrm{avg}} (x)- \mathcal{T} (x)\right]^{2}dx \nonumber\\ &+\beta \int_{0}^{1} \left[ K_{\mathrm{cross}} (x)-\mathcal{T}_{\mathrm{cross}}(x)\right]^{2}dx \nonumber\\ &+ \tan (\theta)\sum_{i}(c_{i}\sigma_{i})^{2}+\lambda  \left[ \sum_{i} c_{i}-f \right] \nonumber \\
&+ \sum_{m=1}^{M_{\mathrm{surf}}} a_m \sum_{i}c_{i}\frac{\psi_{m}(\nu_{i})}{Q_{i}}. \label{eqcostfunction}
\end{align}
There can be $3$ to $5$ terms in the cost function, depending on whether or not a supplementary constraint and/or surface corrections are included. The first two terms are responsible for making the averaging and the cross-term kernels match their respective targets $\mathcal{T}$ and $\mathcal{T}_{\mathrm{cross}}$. The third term of the cost function defines the trade-off between reducing the measurement error bars on the result and improving the match to the target functions. One usually talks of the magnification of the measurement errors. The fourth term is a supplementary constraint on the inversion, usually a unimodularity constraint in the classical SOLA approach. In the following section, we will follow the prescriptions of \citet{Reese} and use a constraint on the sum of the inversion coefficients. The parameters $\beta$ and $\theta$ are trade-off parameters that regulate the balance between different terms in the inversion and $\lambda$ is a Lagrange multiplier. Since the parameters $\beta$ and $\theta$ are free, one can adjust them to modify the results of the inversion but great care has to be taken since they can lead to non-physical results. Finally, the fifth term corrects surface effects in the inversion.
\\
\\
Because of the form of Eq. ($\ref{eqinvresult}$), one has to be careful of the sources of errors on the inverted solution. When the real value of $A_{\mathrm{obs}}$ is known (for example in theoretical analysis), one can nearly always find a set of free parameters so that $A_{\mathrm{inv}}$ will be equal to $A_{\mathrm{obs}}$. However one cannot use the same set of parameters for another inversion and expect the same result. It is therefore necessary to introduce a criterion for which the inversion can be considered as successful and reliable. In this study, we set the parameters by testing several values and choosing the best compromise between reducing the errors and matching the kernels to the target functions. However, the problem is far more complicated since one should analyse how these parameters depend not only on the modes used to carry out the inversion but also on the reference model for every integral quantity. This problem will be discussed in further studies on larger samples to provide relevant results.
\\
\\
The error bars on the inversion result are deduced from the errors bars on the frequency differences, where the errors on individual frequencies are considered to be independent:
\begin{align}
\sigma_{\delta A/A}=\sqrt{\sum_{i}c_{i}^{2}\sigma_{i}^{2}}, \label{eqerrors}
\end{align}
with $\sigma_{i}= \sigma_{\frac{\delta \nu_{i}}{\nu_{i}}}$. However, it is clear that Eq. (\ref{eqerrors}) does not take into account other sources of errors in the inversion, such as non-linear effects in the frequency differences, the mismatch between the averaging or cross-term kernels and their respective target functions, or the errors arising from neglected surface terms in the derivation of the kernels themselves. In other words, the inversion is dependent on the mathematical hypotheses leading to the variational principle \citep{LyndenBell} and other additional simplifications leading to expression (\ref{eqfreqstruc}) \citep[see][]{Gough}. In fact, Eq. (\ref{eqerrors}) only takes into account the amplification of the observational errors, the so-called error magnification, but this is not representative of the accuracy of the method since it does not include all sources of error.
\\
\\
In the test cases of Sect. \ref{gridtest} and \ref{forwardmodel}, the error analysis was performed using the difference between Eq. (\ref{eqrealvalue}) and Eq. (\ref{eqinvresult}), following the method of \citet{Reese}. This leads to the following equation:
\begin{align}
\frac{\delta A-\delta A_{inv}}{A} =& \int_{0}^{1}\left( \mathcal{T}(x)-K_{\mathrm{avg}}(x)\right) \frac{\delta s_{1}}{s_{1}} dx \nonumber \\
&+\int_{0}^{1}\left(\mathcal{T}_{\mathrm{cross}}(x)-K_{\mathrm{cross}}(x)\right)\frac{\delta s_{2}}{s_{2}} dx \nonumber \\
&-\sum_{i}c_{i}\frac{\mathcal{G}(\nu_{i})}{Q_{i}}. \label{eqerrorcontrib}
\end{align}
The first integral is the error contribution originating from the error on the fit of the target to the averaging kernel. We will write it $\sigma_{\mathrm{Avg}}$. The second integral is the error contribution originating from the error on the fit of the target to the cross-term kernel. We will write it $\sigma_{\mathrm{Cross}}$. The third term originates from the surface effects. The above equation does not take into account other sources of error such as the non-linear effects not taken into account in Eq. (\ref{eqfreqstruc}), numerical errors, or the neglect non-adiabatic effects. In what follows, we will lump these errors together with the surface effects and call this $\sigma_{\mathrm{Res}}$ i. e. the ``residual'' errors which are left after having substracted $\sigma_{\mathrm{Avg}}$ and $\sigma_{\mathrm{Cross}}$ from the total error. Of course $\sigma_{\mathrm{Res}}$ can only be obtained in theoretical test cases, where the differences in structural profiles are known beforehand and this specific contribution can be isolated from the kernel contributions.
\subsection{Accuracy and reliability of the solution}
As discussed in the previous section, inversion techniques have to be used with care, especially when modifying the values of the free parameters. First of all, it is necessary to recall that linear inversion techniques are limited to targets, models, and oscillation modes for which Eq. (\ref{eqfreqstruc}) is satisfied to a sufficient accuracy. This means that the reference model already has to be close to the target before the inversion can be computed. Therefore, we propose to make use of the forward modelling method before calculating global characteristics with the inversion technique. For the present study, we used the Optimal Stellar Model (OSM) software developed by R. Samadi (Observatoire de Paris-Meudon) to compute our reference models. We will discuss the fitting process in Sect. \ref{forwardmodel} and further discussions will be made in Sect. \ref{conclusion}.
\\
\\
Once the reference model is obtained to sufficient accuracy, one may carry out the inversion. The free parameters $\beta$ and $\theta$ of the SOLA method can be modified to improve the result. During this optimisation, the contributions from the matching of the averaging kernel, the cross-term kernel and the error magnification must be considered. In fact, one has to make a compromise on the error contributions. One often talks about trade-off between precision and accuracy \cite[see][for a discussion on this problematic in the context of the SOLA method]{Pijpers}. In some of our test cases, we will see that the error magnification can be quite important but on the other hand, having extremely small error bars on an inaccurate result is also unacceptable. 
\section{Inversion procedure for the acoustic radius and the age indicator}\label{acousticage}
\subsection{Definition of the targets and motivations}
As mentioned in the previous section, the first step is to define the global characteristic and its associated target. For this study, we will work with the acoustic radius of the star, denoted $\tau$, and an age indicator, t, based on the integral of the derivative of the sound speed appearing in the asymptotic limit of the small frequency separation. Therefore, the global characteristics we wish to determine are:
\begin{align}
\tau = & \int_{0}^{1}\frac{dx}{c} \label{eqacousticradius},\\
t = & \int_{0}^{1}\frac{1}{x}\frac{dc}{dx}dx. \label{eqageindic}
\end{align}
The acoustic radius will be sensitive to surface effects because of the $1/c$ factor whereas the age indicator will be mostly sensitive to the central regions of the star. During the evolution of the star, the mean molecular weight will grow because of nuclear reactions, leading to a local minimum in the sound speed profile. Therefore its derivative will be very sensitive to the intensity of this minimum and can be related to the age of the star. These targets are also asymptotically related to the large and small frequency separation as follows \citep{Vandakurov, Tassoul}:
\begin{align}
\tau \simeq \frac{1}{2 \Delta \nu}, \\
t \simeq \frac{-4\pi^{2}\nu \tilde{\delta} \nu}{(4 \ell+6)\Delta \nu}, \label{eqfreqcomb}
\end{align}
where we use the symbol $\tilde{\delta}\nu$ to represent the small frequency separation to avoid confusion with the frequency perturbation, $\delta \nu$. It is well known that Eq. (\ref{eqfreqcomb}) is not very accurate for typical solar-like pulsators and that its agreement for models of the sun in its current evolutionary stage is in fact fortuitous \citep{JCDa}. 
\\
\\
It is worth noting that the average large frequency separation is currently the only way to estimate the acoustic radius of a star. This quantity is expected to be sensitive to surface effects like convection and can also be used to characterise structural changes that mimic the evolution of the stellar radius, for example its increase due to the contraction of the core during the evolution of the star. Moreover, the average large separation is also combined with the small frequency separation or other frequency combinations \cite[see][]{JCD93, White} in order to build asteroseismic H-R diagrams. The motivations behind this approach is to estimate the mass and age of the star by using seismic indicators which provide nearly independent information. However asteroseismic diagrams are intrisically limited by two aspects: firstly, the exact relation between frequency separations and the stellar structure is not trivial; secondly, there is only a limited number of different frequency combinations which can be used. In constrast, inversion techniques allow us to target the structural characteristics of our choise based on their relation with stellar properties. Thus, they offer more specific constraints and potentially allow us to distinguish between the different contributions from micro- and macro-physics. 

\subsection{Target for the acoustic radius inversion}
To define the target function of the inversions, we have to calculate the first order relative perturbation of these quantities. For the acoustic radius it is straightforward :
\begin{align}
\frac{\delta \tau}{\tau} & =\frac{1}{\tau}\int_{0}^{1}\frac{-1}{c}\frac{\delta c}{c}dx \nonumber \\
 & =\int_{0}^{1}\frac{-1}{2 \tau c}\frac{\delta c^{2}}{c^{2}}dx.
\end{align}
This result means that the target function is:
\begin{align}
\mathcal{T}_{\tau}=\frac{-1}{2 c \tau}.
\end{align}
Since in this case the perturbation of the acoustic radius is only related to the structural variable $c^{2}$, the contribution of the cross-term kernel has to be suppressed. However, when using the perturbation of $c^{2}$, and the couple $\rho,c^{2}$ in Eq. (\ref{eqfreqstruc}), the cross-term will involve the relative difference in density between the model and the target, potentially leading to high pollution of the solution by the cross-term. It is possible to circumvent this problem by using the structural couple $\rho,\Gamma_{1}$. Indeed, the relative differences on $\Gamma_{1}$ are expected to be small, thereby leading to a smaller cross-term. This can be done by using the following equations: 
\begin{align}
\frac{\delta c^{2}}{c^{2}} &= \frac{\delta \Gamma_{1}}{\Gamma_{1}}+\frac{\delta P}{P}-\frac{\delta \rho}{\rho}, \\
P(x)&=\int_{x}^{1}\frac{m(y)\rho}{y^{2}}dy, \\
m(x) &= \int_{0}^{x}4 \pi x^{2} \rho dx.
\end{align}
Using these equations leads to new target functions defined on the $\rho,\Gamma_{1}$ couple, where we neglected the contribution of the turbulent pressure which is considered as a surface effect:
\begin{align}
\mathcal{T}_{\tau,\mathrm{avg}}=& \frac{1}{2 c \tau}-\frac{m(x)}{x^{2}}\rho \left[\int_{0}^{x}\frac{1}{2 c \tau P}dy \right] \nonumber \\ & - 4 \pi x^{2} \rho \left[ \int_{x}^{1}(\frac{\rho}{y^{2}}\int_{0}^{y}\frac{1}{2 c \tau P}dt) \right]dy. \label{eqtargetavgtau} \\
\mathcal{T}_{\tau,\mathrm{cross}}&=\frac{-1}{2 c \tau}. \label{eqtargetcrosstau}
\end{align}
These definitions can be used directly in Eq. (\ref{eqcostfunction}). Furthermore, we optimise the inversion by defining a supplementary constraint based on homologous relations and extending the method to the non-linear regime by following the approach of \citet{Reese}. 
\\
\\
\subsection{Supplementary constraint and non-linear extension for the acoustic radius}
The idea behind the supplementary constraint is that the result of the inversion should be exact for models which are homologous. In what follows, a procedure satisfying this condition will be described as "unbiased" (not to be confused with the statistical meaning of the word.). To reach this goal, we make use of the knowledge that when using homology, if the density of the model is scaled by a factor $h^{2}$, the frequencies will scale as $h$. By simple analysis of the definition of the acoustic radius, Eq. (\ref{eqacousticradius}), we can see that it will scale as the inverse of the frequencies, $1/h$. Therefore, to the first order, the relative variation of the acoustic radius should be the opposite of the relative variation of the frequencies. This means that if $\delta \nu / \nu = \epsilon$, then $\delta \tau / \tau = -\epsilon$. Furthermore, we know that for linear inversion techniques, the inverted correction is obtained from a linear combination of relative frequency differences. Therefore, if the sum of the coefficient is equal to $-1$, the inverted correction will be exact for models in a homologous relation. 
\\
\\
The non-linear extension is based on an iterative process involving successive scalings of the model to reach an optimal point for which there is no further correction by the inversion technique. We will see after some developments that this process can be by-passed and that the solution can be obtained directly. However, to grasp the philosophy of this extension, it is easier to see it first as an iterative process. First, we carry out an inversion of the acoustic radius for a first reference model with a given $\tau_{ref}$ and obtain a new estimate of the acoustic radius $\tau_{inv,0}$. We now define a scale factor $q_{0}=\frac{\tau_{inv_{0}}}{\tau_{ref}}$, used to scale the reference model, bringing it closer to the observed target. We can use this scaled model as a reference model for which another inversion can be carried out. Indeed, the frequencies have been scaled by the factor $h_{0} = \frac{1}{q_{0}}$ and the relative differences between the frequencies of the scaled reference model and those of the target are now given by:
\begin{align}
\frac{\nu_{\mathrm{obs}}-h_{0}\nu_{\mathrm{ref}}}{h_{0}\nu_{\mathrm{ref}}}=\frac{1}{h_{0}} \left( \frac{\delta \nu}{\nu}+1 \right) -1,
\label{eqnonlinearscaling}
\end{align}
where $\nu_{\mathrm{obs}}$ is the observed frequency and $\nu_{\mathrm{ref}}$ the frequency of the unscaled reference model. Now for the $j^{th}$ iteration, the inverted acoustic radius can be expressed as follows:
\begin{align}
\tau_{\mathrm{inv,j+1}}&=\frac{\tau_{ref}}{h_{j}}\left[1+\sum_{i} c_{i}\left[ \frac{\nu_{\mathrm{obs},i}-h_{j}\nu_{\mathrm{ref},i}}{h_{j} \nu_{\mathrm{ref},i}} \right] \right] \nonumber \\
& = \frac{\tau_{\mathrm{ref}}}{h_{j}} \left[ 1+\sum_{i}c_{i}\left[ \frac{1}{h_{j}}(\frac{\delta \nu_{i}}{\nu_{i}}+1)-1 \right] \right] \nonumber \\
&=\tau_{\mathrm{ref}}\left[\frac{2}{h_{j}}+\left[\frac{1}{h^{2}_{j}}\left( \sum_{i}c_{i}\frac{\delta \nu_{i}}{\nu_{i}}-1\right)\right]\right], \label{eqnonlinearparabola}
\end{align}
where we have also used the fact that the sum of the inversion coefficient is $-1$ for an ``unbiased'' acoustic radius inversion. Now we also have that $\tau_{inv,j+1}=\frac{\tau_{ref}}{h_{j+1}}$, by definition of our iterative process. Using this definition and rewriting Eq. (\ref{eqnonlinearparabola}) in function of $q_{j}$ and $q_{j+1}$, we obtain the following expression:
\begin{align}
q_{j+1}= 2 q_{j}+q^{2}_{j}\left[\sum_{i}c_{i}\frac{\delta \nu_{i}}{\nu_{i}}-1 \right]=f(q_{j}). \label{eqfixedpoint}
\end{align}
where we have introduced the function, $f$. If the above iterations converge, then the limit, $q_{\mathrm{opt}}$, will be a fixed point of $f$, i.e. $f(q_{\mathrm{opt}}) = q_{\mathrm{opt}}$. Convergence is guaranteed over a neighbourhood around $q_{\mathrm{opt}}$ provided $\vert f^{'}(q_{\mathrm{opt}}) \vert < 1$.  Given the simplicity of $f$, we choose to bypass the iterative method by solving directly $f(q) = q$. There are two solutions. The first is $q = 0$. However, it leads to an unphysical result, and would tend not to be the result of an iterative process since $f^{'}(0) = 2$. The second solution is the one we're searching for:
\begin{align}
q_{\mathrm{opt}}=\frac{-1}{\sum_{i}c_{i}\frac{\nu_{\mathrm{obs},i}}{\nu_{\mathrm{ref},i}}} \label{optimaltau},
\end{align}
Furthermore, it turns out that $f^{'}(q_{\mathrm{opt}}) = 0$. Hence, had we applied an iterative method, the convergence would have been quadratic. The associated acoustic radius is $\tau_{\mathrm{inv}} = q_{\mathrm{opt}} \tau_{\mathrm{ref}}$. However, one must be aware that the error bars given by Eq. (\ref{eqerrors}) on the final result are modified as follows if we assume that $\sigma_{i} \ll 1$ and that the errors on the individual frequencies are independent(see Appendix \ref{secapperror} for the demonstration of this formula):
\begin{align}
\sigma_{\tau_{\mathrm{min}}}=q_{\mathrm{opt}}^{2}\tau_{\mathrm{ref}}\sqrt{\sum_{i}c_{i}^{2}\sigma_{i}^{2}}.
\label{errornonlinear}
\end{align}
\subsection{Target for the age indicator inversion}\label{targetage}
By considering the perturbation of Eq. (\ref{eqageindic}), we obtain the following target:
\begin{align}
\frac{\delta t}{t}&=\frac{1}{t}\int_{0}^ {1}\frac{1}{x}\frac{d \delta c}{dx}dx \nonumber\\
&=\frac{1}{t}\int_{0}^ {1}\frac{1}{x}\frac{dc}{dx}\frac{\frac{d \delta c}{dx}}{\frac{d c}{dx}}dx.
\end{align}
The fact that we divide and multiply by the sound speed derivative is simply due to the fact that the kernels will be unable to match the function $1/x$ in the centre. Therefore we use this operation to define an easier target for the inversion and express the problem in terms of the relative perturbation of the sound speed derivative. The target is then given by:
\begin{align}
\mathcal{T}_{t}(x)=\frac{\frac{1}{x}\frac{dc}{dx}}{\int_{0}^{1}\frac{1}{x}\frac{dc}{dx} dx}.
\end{align}
If we now consider Eq. (\ref{eqfreqstruc}), we can use an integration by parts to obtain inversion kernels in terms of the sound speed derivative:
\begin{align}
\int_{0}^{1}K^{n,\ell}_{c^{2},\rho}\frac{\delta c^{2}}{c^{2}}dx=& - \int_{0}^{1} \left( \int_{0}^{x}\frac{2K^{n,\ell}_{c^{2},\rho}}{c}dy \right) \frac{dc}{dx}\frac{\frac{d \delta c}{dx}}{\frac{d c}{dx}}dx  \nonumber \\
& + \left[ \left( \int_{0}^{x} 2 \frac{K^{n,\ell}_{c^{2},\rho}}{c} ds \right) \delta c \right]^{1}_{0}.
\end{align}
In the second term of this expression, the central evaluation is exactly $0$ because the kernels are proportional to $x^{2}$ and the surface evaluation has been neglected because numerical tests have shown that its amplitude was $60$ to $150$ times smaller than the first term for modes with higher degree and radial order, and even smaller for lower degree and radial order modes. We then define the structural kernels for the sound speed derivative as follows:
\begin{align}
K^{n,\ell}_{dc/dx,\rho}=-\frac{dc}{dx}\int_{0}^{x}\frac{2K^{n,\ell}_{c^{2},\rho}}{c}dy.
\end{align}
By identification, we also obtain that $K^{n,\ell}_{\rho,dc/dx}=K^{n,\ell}_{\rho,c^{2}}$, which will be associated with the cross-term kernel. When deriving the targets for the acoustic radius, it was rather straightforward to obtain the cost-function for the inversion. In the case of the age indicator, we will show in Sect. \ref{gridtest} that the cost function defined in Eq. (\ref{eqcostfunction}) is not adequate. Therefore we defined a new way to carry out a SOLA inversion: trying to match the antiderivative of the averaging kernel with the antiderivative of the target function. This modification is motivated by the oscillatory behaviour of the structural kernels which is unsuitable for the age indicator inversion. Using this method, the cost function is defined as follows:
\begin{align}
J_{t}=&\int_{0}^{1}\left[ \int_{0}^{x}\mathcal{T}(y)dy-\int_{0}^{x}K_{\mathrm{avg}}(y)dy \right]^{2}dx \nonumber \\ & + \beta \int_{0}^{1}K_{\mathrm{cross}}^{2}(x)dx+\tan(\theta)\sum_{i}(c_{i} \sigma_{i})^{2}\nonumber \\ &+\lambda \left[ \sum_{i} c_{i}-f \right].
\end{align}
The fourth term contains the supplementary constraint we will define in the next section, and once again we do not consider the ad-hoc surface correction term. As for the acoustic radius, we can determine the value of the number $f$ using homologous relations and add a non-linear extension to the method. 
\subsection{Supplementary constraint and non-linear extension for the age indicator}
The supplementary constraint is obtained in the same way as for the acoustic radius inversion. We know that the frequencies scale with $\sqrt{\upsilon/\epsilon^{3}}$ for a scale factor of $\upsilon$ in mass and $\epsilon$ in radius, or in other terms a scaling factor $\upsilon/\epsilon^{3}$ in density. It is easy to show that the adiabatic sound speed will scale as $\sqrt{\upsilon/ \epsilon}$ and therefore its derivative will scale as $\sqrt{\upsilon/\epsilon^{3}}$. This means that the first order relative correction of the age indicator has to be the same as the frequency correction for models in a homologous relation. Again, we can find a constraint on inversion coefficients so that the inverted correction will be exact for models in a homologous relation. In this case, it means that the sum of the inversion coefficients needs to be equal to $1$ to ensure that the correction will be the same for both frequencies and $t$.
\\
\\
It is also possible to try to extend this inversion to the non-linear regime using the iterative method of Eq. (\ref{eqnonlinearscaling}). Using this definition and the constraint on the sum of the inversion coefficients, we obtain:
\begin{align}
t_{\mathrm{inv}}&=ht_{\mathrm{ref}}\left[ 1+\sum_{i}c_{i} \left[\frac{1}{h}(\frac{\delta \nu_{i}}{\nu_{i}}+1)-1 \right] \right] \nonumber \\
&= t_{\mathrm{ref}}(1+\sum_{i}c_{i}\frac{\delta \nu_{i}}{\nu_{i}}).
\end{align}
We now see that the inverted result is independent of the scaling factor $h$ meaning that the effect of the iterative process described for the acoustic radius is already included in the linear method. However, this does not mean that the SOLA method is non-linear, nor that a non-linear inversion could not be defined by some other approach. 
\subsection{Comparison with asymptotic laws based on frequency separations}
In the next sections, we will compare the results of SOLA inversions to other techniques based on frequency separations. We stress that these methods are not inversion techniques; we simply express asymptotic laws in a differential formulation to relate them to a linear combination of frequency differences. 
\\
\\
It was shown by \citet{Vandakurov} that the average large frequency separation is asymptotically related to the acoustic radius in the following way:
\begin{align}
\tau \approx \frac{1}{2\left< \Delta \nu \right>}. \label{eqlargesep}
\end{align}
When we linearise this relation we obtain:
\begin{align}
\frac{\delta \tau}{\tau}\approx-\frac{\delta \left< \Delta \nu \right>}{\left< \Delta \nu \right>}=\sum_{i}c_{i}\frac{\delta \nu_{i}}{\nu_{i}} \label{eqtaulargesep},
\end{align}
where we used the fact that the average large separation is simply a linear combination of frequencies to derive coefficients $c_{i}$. In much the same way as was done for inversion coefficients. These coefficients can be inserted into Eqs. (\ref{eqaveraging}) and (\ref{eqcrossterm}) in order to obtain averaging and cross-term kernels for this method. These kernels can then be directly compared with those coming from the SOLA inversion technique, thereby allowing a quantitative comparison of the two methods. In our study, the average large separation was determined by a $\chi^{2}$ fit \citep{Kjeldsen}.  If we apply the non-linear extension to the above relation, we obtain the following result:
\begin{align}
\tau_{\mathrm{inv}}=-\frac{\tau_{\mathrm{ref}}}{\sum_{i}c_{i}\frac{\delta \nu_{i}}{\nu_{i}}-1}=\frac{\tau_{\mathrm{ref}}}{(\frac{\left<\Delta \nu\right>_{\mathrm{obs}}}{\left<\Delta \nu \right>_{\mathrm{ref}}})}=\frac{\gamma_{\tau}}{\left< \Delta \nu \right>_{\mathrm{obs}}}.
\label{eqtauscaling}
\end{align}
where $\gamma_{\tau}=\tau_{\mathrm{ref}}\left<\Delta\nu\right>_{\mathrm{ref}}$. Although Eq. (\ref{eqtauscaling}) is very similar to Eq. (\ref{eqlargesep}), there are some subtle, yet important, differences.  Indeed, the proportionality constant $\gamma_{\tau}$ is not, in general equal to $1/2$ (as given by the original asymptotic formula), but has been specifically adapted to the reference model for that particular range of modes.  Likewise, SOLA inversions are calibrated on the reference model, but they also go a step further by optimising the frequency combination so as to be as sensitive as possible to the acoustic radius.
\\
\\
We now turn our attention to the age indicator and the small frequency separation. We know from \citet{Tassoul} that the small frequency separation is asymptotically and approximately related to the derivative of the sound speed by the following relation:
\begin{align}
\tilde{\delta} \nu \approx \frac{-(4 \ell +6) \Delta \nu}{4 \pi^{2} \nu_{n, \ell}}\int_{0}^{R}\frac{dc}{dr}\frac{dr}{r} \label{eqsmallsep},
\end{align}
which can be reformulated in the form of Eq. (\ref{eqfreqcomb}). The relative perturbation of this equation will be a frequency combination, thereby allowing us to write:
\begin{align}
\frac{\delta \frac{\nu \tilde{\delta} \nu}{\Delta \nu}}{\frac{\nu \tilde{\delta \nu}}{\Delta \nu}}&= \sum_{i}c_{i}\frac{\delta \nu_{i}}{\nu_{i}} \nonumber\\
&\approx \frac{\delta t}{t}.
\end{align}
In other words, by using the relative perturbation of Eq. (\ref{eqfreqcomb}), we can define inversion coefficients leading to the following estimate of the indicator $t$ :
\begin{align}
t_{\mathrm{inv}}&= t_{\mathrm{ref}} \left( 1+\sum_{i}c_{i}\frac{\delta \nu_{i}}{\nu_{i}} \right) \nonumber \\
       &= \frac{t_{\mathrm{ref}} \left(\frac{\nu \tilde{\delta \nu}}{\Delta \nu} \right)_{\mathrm{obs}}}{ \left( \frac{\nu \tilde{\delta \nu}}{\Delta \nu}\right) _{\mathrm{ref}}} = \gamma_{t} \left( \frac{\nu \tilde{\delta \nu}}{\Delta \nu} \right) _{\mathrm{obs}}.
\end{align}
Again we find a proportionality constant $\gamma_{t}$ adapted to the reference model and the observed modes. Using Eq. \ref{eqsmallsep}, one would find $\gamma_{t}=\left<-4\pi^{2}/(4  \ell +6)\right>$. We will see in the next sections that the indicators determined by directly applying the asymptotic relations are inaccurate compared to the SOLA method and the estimates defined in this section. In \citet{Reese}, the same technique is also applied to the scaling relationship between the mean density and the large frequency separation, and to another technique which includes the empirical surface corrections of \citet{Kjeldsen}. In Sect. \ref{forwardmodel}, we will compare the three above procedures for estimating the mean density. Following the notations of \citet{Reese}, we will refer to Kjeldsen et al.'s approach as the KBCD method\footnote{Eq. ($26$) in \citet{Reese}.}. The methods presented in this paper are summarised in Table \ref{methods}.
\begin{table*}[t]
\caption{Methods used for the determination of $t$, $\tau$ and $\bar{\rho}$.}
		\label{methods}
  \centering
\begin{tabular}{c | c | c }
\hline \hline
$\bar{\rho}$ determination& $t$ determination& $\tau$ determination\\ \hline 
SOLA with $\theta=10^{-2}$, $\beta=10^{-6}$ & SOLA with $\theta=10^{-8}$, $\beta=10^{-2}$& SOLA with $\theta=10^{-2}$, $\beta=10^{-6}$ \\ 
$\left< \Delta \nu \right>$ estimate &$\left< \tilde{\delta} \nu \right>$ estimate& $\left< \Delta \nu \right>$ estimate\\ 
KBCD estimate with $b=4.9$& $-$ & $-$  \\ \hline
\end{tabular}
\end{table*}
\section{Test case with a grid of model}\label{gridtest}
\subsection{Targets and grid properties}
The first test carried out used the model grid and the targets of \citet{Reese}. The goal of this test was to determine the reliability of the inversion when no forward modelling\footnote{Strictly speaking, the term ``forward modelling'' refers to solving the direct problem (see e.g. \citet{Tarantola}, Sect. $1.3$), i.e. predicting the results (or in our case the pulsation frequencies) for a given model.  However,  in the asteroseismic literature \citep[see e. g.][]{Charpinet}, the term ``forward modelling'' has also come to mean ``execution of the forward problem using [stellar] models with a few adjustable parameters, and the calibration of those parameters by fitting theory to observations'' \citep{Gough85}.  In what follows, we use this latter definition.} was performed. The model grid consists of $93$ main sequence and pre-main sequence models with masses ranging from $0.8M_{\odot}$ to $0.92$ $M_{\odot}$ and ages ranging from $28$ Myr to $17.6$ Gyr. These models were downloaded from the CoRoT-HELAS website and additional information on their physical characteristics can be found in \citet{Marques} and \citet{Reese}. 
\\
\\
In this paper we will only present the results for two of the three targets, models $A'$ and $B$, following the naming convention of \citet{Reese}. The characteristics of these targets are summarised in Table \ref{tabcaractarg}. The results for the first target were similar to those for $B$ so we do not present them here. Model $A'$ is in fact the first target of \citet{Reese}, denoted model $A$ in their study, to which has been added an ad-hoc $50 \%$ increase of the density in the surface regions in the form of a hyperbolic tangent. Model $B$ is radically different from the models of the grid since it includes rotational mixing, diffusion and follows the solar mixture of \citet{Asplund} rather than that of \citet{GrevNoels}, as used in the grid. We used a set of 33 oscillation modes ranging from $\ell=0$ to $\ell=2$ and from $n=15$ to $n=25$. The error bars on the observed frequencies were set to $0.3$  $\mu Hz$.
\begin{table}[t]
\caption{Characteristics of targets $A'$ and $B$.}
		\label{tabcaractarg}
  \centering
\begin{tabular}{r | c | c }
\hline \hline
 & \textbf{Model $A'$}& \textbf{Model $B$}\\ \hline 
$\tau$ ($\mathrm{s}$) & $2822.07$& $2823.53$ \\ 
$t$ ($\mathrm{s^{-1}}$) &$-2.640\times 10^{-3}$& $-2.534\times 10^{-3}$\\ 
Mass ($\mathrm{M_{\odot}}$)& $0.9$ & $0.92$  \\
Radius ($\mathrm{R_{\odot}}$) & $0.821$ & $0.825$ \\ 
Age ($\mathrm{Gyr}$) &$1.492$ &$2.231$  \\ 
$T_{\mathrm{eff}}$ ($\mathrm{K}$) & $5291$ & $5291$ \\ 
$\log(g)$ ($\mathrm{dex}$) & $4.563$ & $4.569$ \\ \hline
\end{tabular}
\end{table}
\subsection{Results for the acoustic radius}\label{secresulttau}
\begin{figure*}[t]
	\centering
		\includegraphics[width=18.7cm]{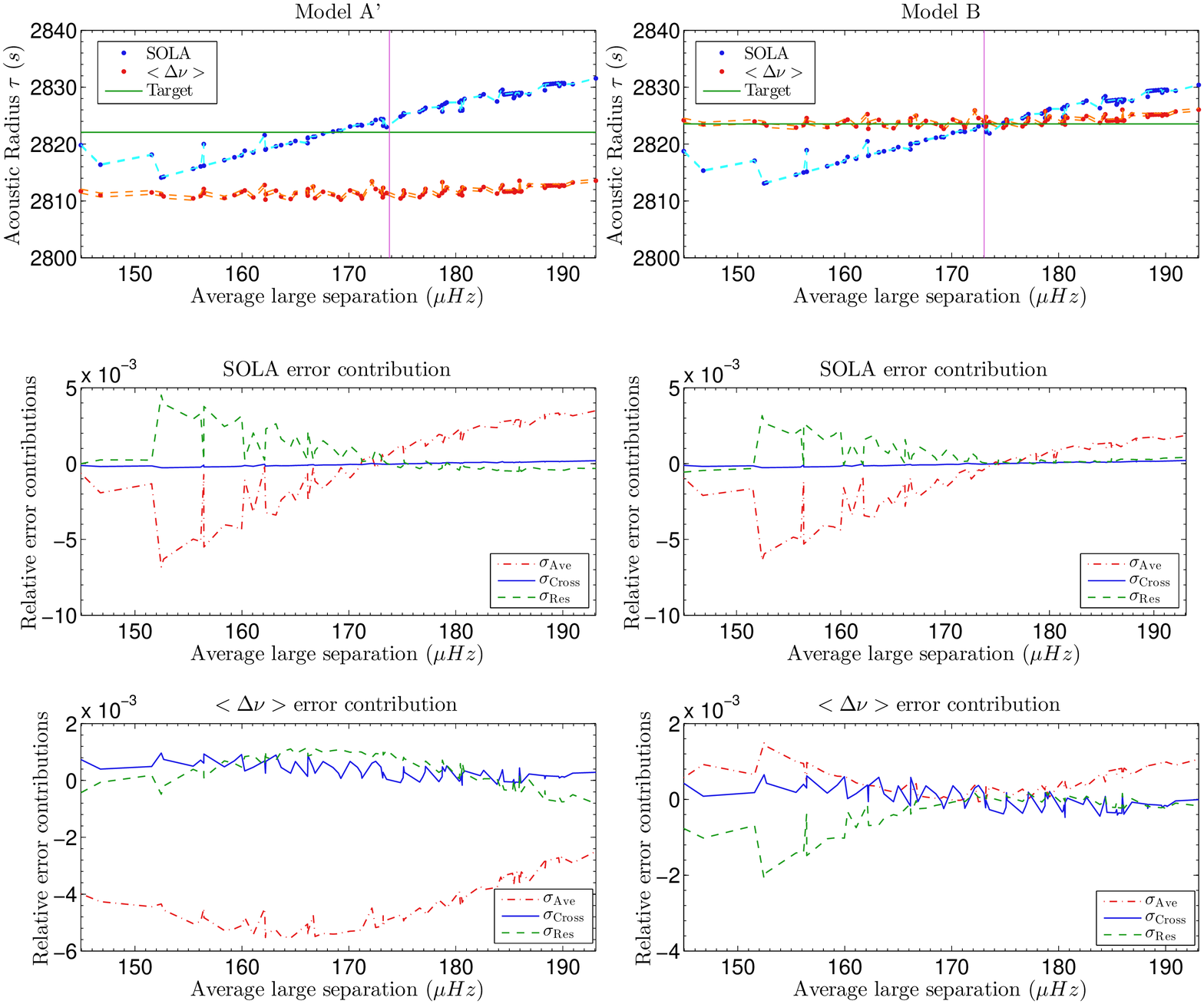}
	\caption{The left-hand panels show inversion results for model $A'$, whereas the right-hand side is for model $B$. The top panels show SOLA inversion results (blue) and estimates based on the large frequency separation (red). The figures below show the different error from Eq. (\ref{eqerrorcontrib}) terms which appear in the SOLA inversions (middle panels) and the large separation (lower panels). The results and error contributions are given for every model of the grid such that the abscissa of these figures is the average large separation of each reference model.}
		\label{figresultau}
\end{figure*}

\begin{figure*}[t]
	\centering
		\includegraphics[width=18cm]{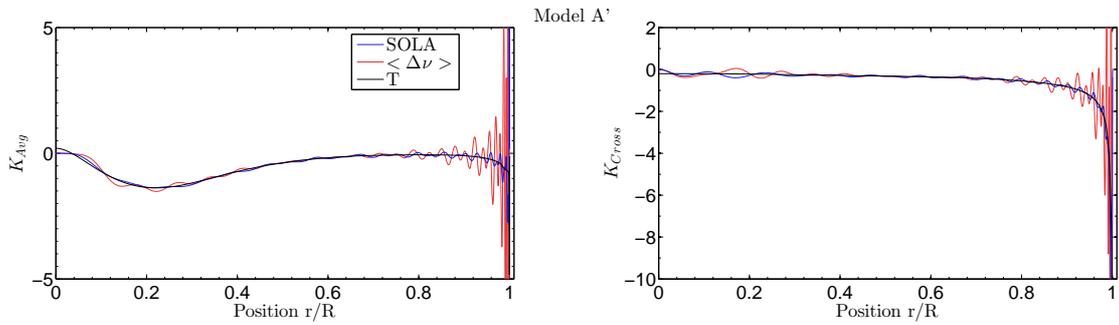}
\begin{center}	
	\caption{Averaging and cross-term kernels for the inversion of the acoustic radius for model $A'$. The target function is represented in black, the results for the $\Delta \nu$ relation in red, and those from SOLA inversions in blue.}
			\label{figkerneltau}
	\end{center}
\end{figure*}

The results for the acoustic radius for models $A'$ and $B$ are represented in Fig. \ref{figresultau}. 
The values of the parameters $\theta$ and $\beta$ are chosen so as to improve the match between the averaging and cross-term kernels, and their respective targets. The optimal values are $\theta=10^{-2}$ and $\beta=10^{-6}$. The small value of $\beta$ is due to the fact that the second target ($\mathcal{T}_{\mathrm{cross}}$ defined by Eq. (\ref{eqtargetcrosstau})) will be multiplied by the corrective term $\delta \Gamma_{1}/\Gamma_{1}$ which is rather small. Likewise, $\theta$ could be reduced because the error bars were not dramatically affected by changes in the value of this parameter. Because the structure of the target is known, it is possible to plot all error contributions to the inversion results as in Eq. (\ref{eqerrorcontrib}) and the error analysis described at the end of Sect. \ref{generalapproach}. These contributions are represented for targets $A'$ and $B$ in Fig.\ref{figresultau} and the kernels for model $A'$ are represented in Fig. \ref{figkerneltau}.
We see that the cross-term is not responsible for the errors of the SOLA inversions and that the matching of the averaging kernel is the leading error term. Also, we observe sometimes a compensation of the residual error and the averaging kernel error for the SOLA method and that the correction based on the large frequency separation can have smaller errors than SOLA, despite its oscillatory behaviour. The value of the least square fits of the kernels for model $A^{'}$ for the $\tau$ and $t$ inversions are illustrated on figure \ref{figKerFits} where we compute the squared difference between the kernel and its target for each reference model of the grid. However, it should be noted that these errors tend to compensate and that this compensation is the reason for the slightly more accurate results for model $B$, as can be seen on the right hand side of the figure. Such compensations have also been observed for mean density inversions, but in the case of model $A^{'}$ and other test cases, they did not occur, as can be seen in the error plots in Fig. \ref{figresultau}. Thus, this technique is unable to account for surface effects and its reliability for observed stars is questionable. If we use directly the asymptotic relation for the acoustic radius, i.e. if we apply Eq. (\ref{eqlargesep}), we obtain $\tau=2691$ s for model $A'$ and $ \tau=2890$ s for Model $B$ which is even less accurate than both SOLA inversions and the improved $\Delta \nu$ approach.
\subsection{Results for the age indicator}
\begin{figure*}[t]
	\centering
		\includegraphics[width=17.5cm]{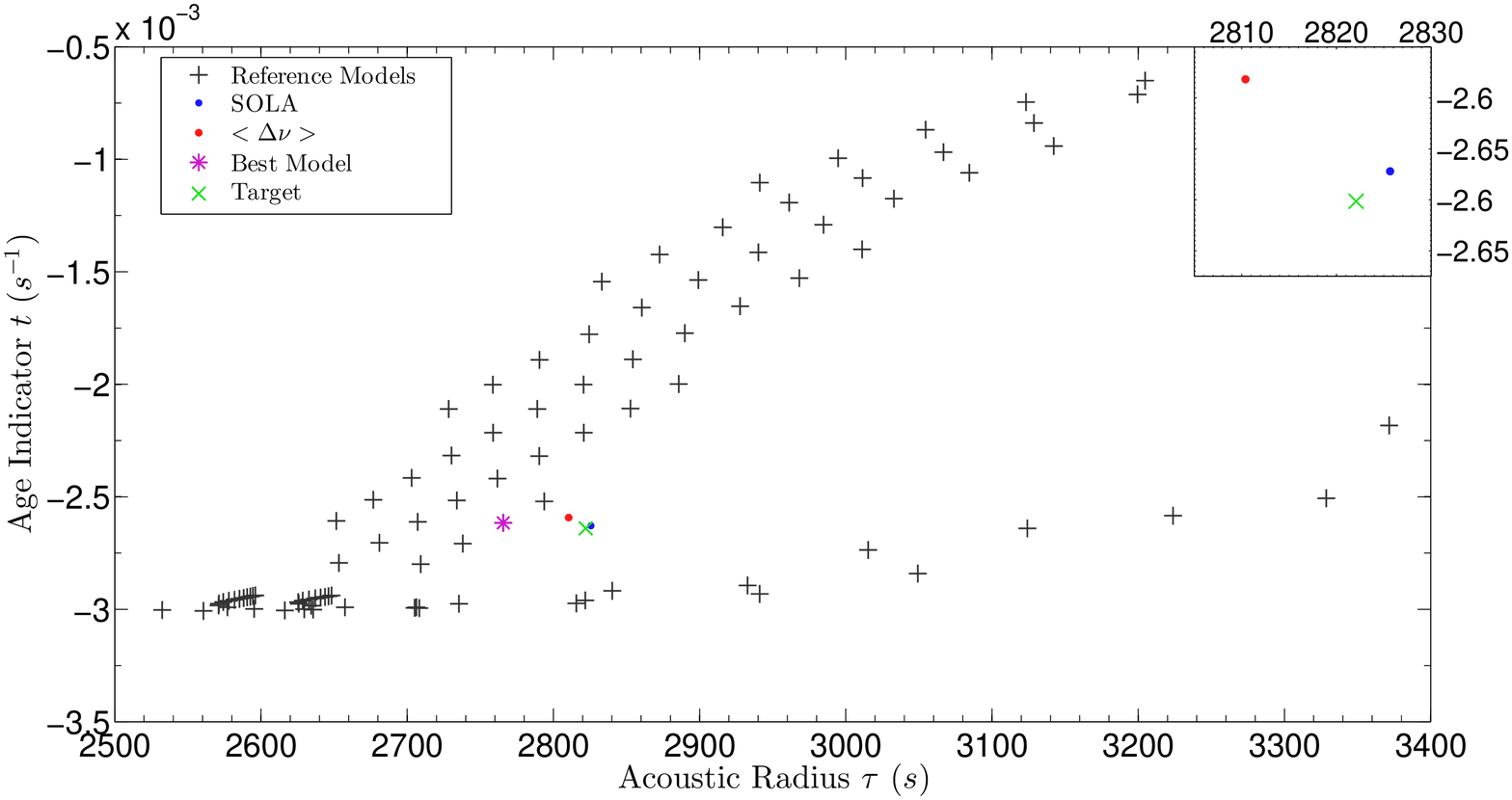}
	\caption{Inversion results for the model with the best small frequency separation. In the main part of the figure, the grid models are represented by the black $+$, the best model is the purple $*$, model $A'$ the green $X$, the SOLA result is in blue and the large frequency separation result in red. The inset shows an enlarged view of the region around model $A'$. }
		\label{figbestmodelchoice}
\end{figure*}

The results of the age indicator inversions for models $A'$ and $B$ are the same, thus we only present them for model $A'$.
\begin{figure*}[t]
	\centering
		\includegraphics[width=18.7cm]{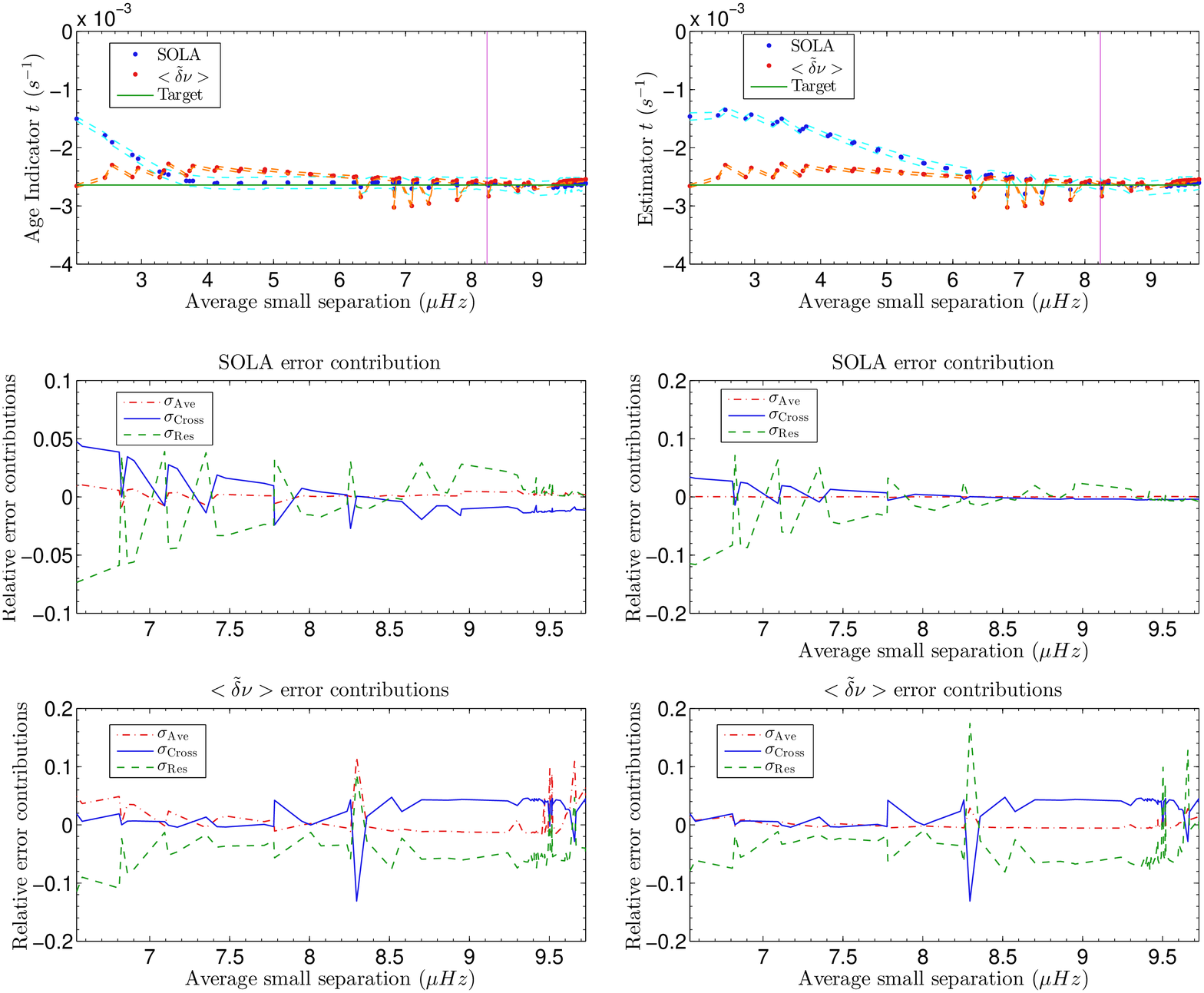}
	\caption{Inversion results for the age indicator and estimates based on the small frequency separation, using the grid of models.  The left column shows the results in which the averaging kernels are optimised, whereas in the right column, it is the antiderivative of the averaging kernels which are optimised.  The top two panels show the inversion results, the middle two panels show the errors from Eq. (\ref{eqerrorcontrib}) in the SOLA inversions, and the bottom two panels are the errors from Eq. (\ref{eqerrorcontrib}) from the improved small frequency separation technique. The results and error contributions are given for every model of the grid such that the abscissa of these figures is the average large separation of each reference model.}
		\label{figresultsmallsep}
\end{figure*}  
They show the limit of our inversion techniques when there is no criterion to choose the reference model. From Fig. \ref{figresultsmallsep}, it is clear that the SOLA inversion technique failed to reproduce the results for a subgrid of models. This is simply due to the large range of ages of the reference models. One has to recall that the SOLA approach is based on the integral Eq. (\ref{eqfreqstruc}), which itself is based on the variational principle, only valid for small perturbations. The error plot also shows that SOLA inversions benefit from error compensations, which is problematic for observed stars. The second problem is that when plotting the averaging and cross-term kernels, we see that the results are rather poor. The parameters for these inversions were: $\theta=10^{-6}$ and $\beta=10^{-4}$.
\begin{figure*}[t]
	\centering
		\includegraphics[width=17.5cm]{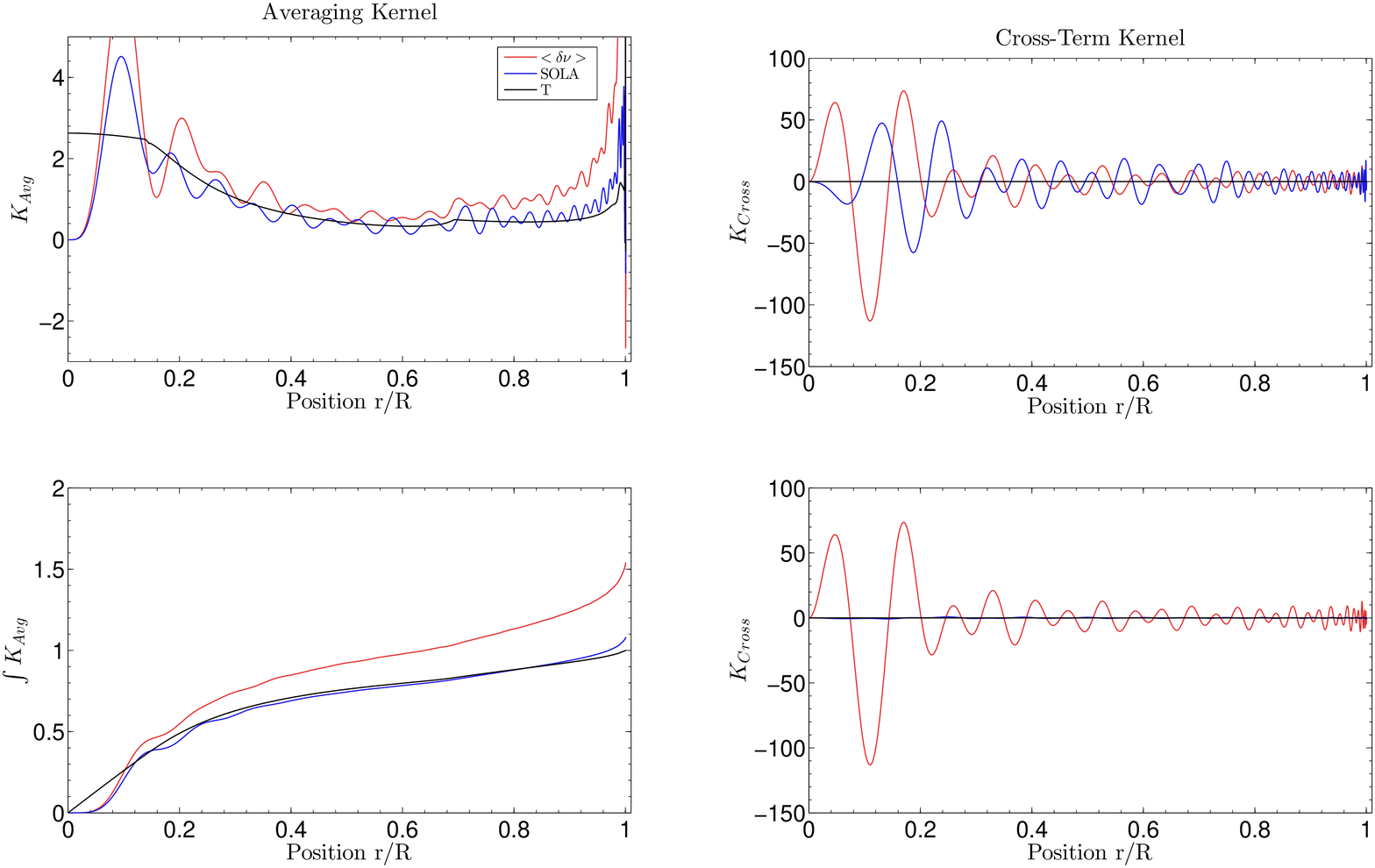}
	\caption{The upper panels illustrate the averaging and cross-term kernels for the model with the best small frequency separation by optimising on the averaging kernel itself. The lower panels illustrate the same results by optimising the antiderivative of the averaging kernel. The target function is in black, the results for the small frequency separation estimate in red and those for SOLA inversions in blue.}
		\label{figsmallsepkernels}
\end{figure*} 
When carrying out an inversion on an observed star, one can only assess the quality of the inversion based on how well the averaging and the cross-term kernels fit their respective target functions. Therefore being able to obtain accurate results is not sufficient: the accuracy must be related to the quality of the fit of the targets, otherwise one would never be able to determine if the inversion was successful or not. Figure \ref{figsmallsepkernels} illustrates the exact opposite for both our techniques. Therefore, we modified the age indicator inversion by using the antiderivative of the target function rather than the target itself, as described in Sect. \ref{targetage}. We then see that the inversion failed on a larger subgrid than before, but this failure is inevitable because of the properties of the reference grid. The set of parameters for these inversions was $\theta=10^{-8}$ and $\beta=10^{-2}$. The parameter $\beta$ was increased to annihilate the effect of the cross-term and $\theta$ was reduced thanks to its small impact on the error bars. However, we need to define a criterion to select a model for which the result is reliable. We simply take the model with the closest average small frequency separation to the target. The results for this choice are illustrated in Fig. \ref{figbestmodelchoice}. In this case, it is clear that the SOLA inversion is superior to the estimate based on the small frequency separation and this leads to the definition of a new framework in which to carry out inversions for this indicator more accurately.
\section{Test case for targets using forward modelling} \label{forwardmodel}
\subsection{Definition of the framework}
In the previous section, we saw that by simply choosing the best model of the grid in terms of the arithmetic average of the small frequency separation, we could achieve very accurate results. However, the validity of Eq. (\ref{eqfreqstruc}) for the ``best model'' of the grid and the target is still questionable and one could wish to achieve an even greater accuracy. Using forward modelling of the target is the best way to obtain a model that is sufficiently close to enable the use of the variational principle, thereby leading to successful linear inversions. We choose the software Optimal Stellar Model (OSM)\footnote{The OSM software can be downloaded from \url{https://pypi.python.org/pypi/osm/}}, developed by Réza Samadi, to carry out the forward modelling using the arithmetic average of the large separation and the small frequency separations of the observed frequency set as seismic constraints and the mass and age of the reference model as free parameters. This optimization strategy is purely arbitrary and further studies will be needed to determine how other approaches can be used. However, regardless of what quantities (e.g. individual small separations or other seismic indicators) and analysis methods (e.g. MCMC algorithms or genetic algorithms) are used to select the reference model, the inversion will be carried out afterwards, since it is able to depart from the physical assumptions used by the stellar evolution code when constructing the reference model. To ensure that differences still remain between our reference model and our targets, we deliberately use different values for the metallicity or the mixing-length parameter, add turbulent pressure to the target, or use non-adiabatic computations for the ``observed" frequencies. Therefore the forward modelling process will always intentionally be unable to reproduce the target within an accuracy that would make the inversion step useless. The tests were carried out using the CESTAM evolutionary code \citep{MarquesCestam}, and the  Adipls \citep{adipls}, the LOSC \citep{Scuflaire} and MAD pulsation \citep{Dupret, Dupretconv} codes. We used the same modes as for the model grid tests, namely with $\ell$ ranging from $0$ to $2$ and $n$ ranging from $15$ to $25$. The error bars on the frequencies were set to $0.33$ $\mu Hz$. We will compare the results from the SOLA method with those from improved estimates based on the average large separation as in the previous section. One could ask why we are not using the arithmetic average of the large separation to carry out the correction. In fact this quantity is already fitted to within $0.2$ $\mu Hz$ of its target value with the forward modelling process, and cannot therefore be improved upon. Concerning the values of the $\theta$ and $\beta$ parameters, we keep the same values as in the previous section, i.e. $\theta=10^ {-2}$ and $\beta=10^{-6}$ for the acoustic radius and the mean density, $\theta=10^ {-8}$ and $\beta=10^{-2}$ for the age indicator.
\subsection{Test case with different metallicity and $\alpha_{\mathrm{MLT}}$}
The first test made use of a $0.95$ $M_{\odot}$ and a $1.05$ $M_{\odot}$ model, denoted targets $1$ and $2$, respectively. The characteristics of the targets are summarised in the Table \ref{tabcaractargalphaZ}. The first step was to carry out the forward modelling of these targets with the OSM software using the fixed parameters $Z=0.0135$ and $\alpha_{\mathrm{MLT}}=1.522$ for the reference models. In tables \ref{tabresultsMetal} and \ref{tabresultsAlpha}, we summarise the inversion results with their error bars for both models. We can see from this table that the error bars are underestimated for the acoustic radius and mean density inversions. This results from the definition (\ref{eqerrors}) which only accounts for the propagation of observational errors but neglects the contributions related to the inversion process itself or to the validity of Eq. (\ref{eqfreqstruc}). However, the error bars from the age indicator are more important. We stress that quantifying errors of inversion techniques is still problematic and require further theoretical studies. We also analysed the different contributions $\sigma_{i}$ and found that compensation was present to a lesser extent in SOLA inversions than in the other correction techniques. This is a direct consequence of the quality of the kernel fits with SOLA.
\\
\\
We also observed that the cross-term kernel contribution could sometimes be rather important in the mean density and acoustic radius inversions.
\begin{table}[t]
\caption{Characteristics of targets $1$ and $2$.}
\label{tabcaractargalphaZ}
  \centering
\begin{tabular}{r | c | c }
\hline \hline
 & \textbf{Model $1$}& \textbf{Model $2$}\\ \hline
\textit{Mass ($\mathrm{M_{\odot}}$)}& $0.95$ & $1.05$  \\
\textit{Radius ($\mathrm{R_{\odot}}$)} & $0.868$ & $0.988$ \\ 
\textit{Age ($\mathrm{Gyr}$)} &$1.8$ &$1.5$  \\ 
\textit{$T_{\mathrm{eff}}$ ($\mathrm{K}$)} & $5284$ & $5912$ \\ 
\textit{$\log(g)$ ($\mathrm{dex}$)} & $4.538$ & $4.469$ \\
\textit{$Z$} & $0.015$ & $0.0135$ \\
\textit{$\alpha_{\mathrm{MLT}}$} & $1.522$ & $1.7$ \\
\hline
\end{tabular}
\end{table}
\begin{figure*}[t]
    \begin{center}
    \textbf{Kernels for $\tau$, $\bar{\rho}$ and $t$ for Model $1$}
    \end{center}
	\centering
		\includegraphics[width=17.5cm]{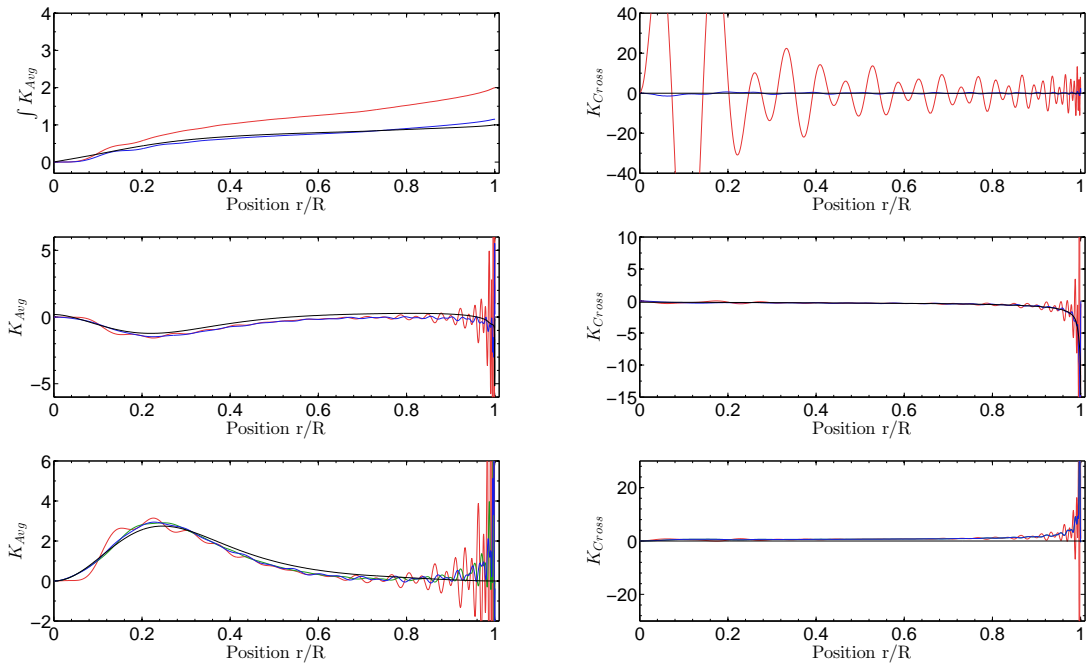}
	\caption{(Colour online) kernels for the test case with different metallicity. Averaging kernels (Left) and cross-term kernels (Right) for the age indicator inversion (top panels), the acoustic radius inversion (middle panels) and the mean density (lower panels). The SOLA method is in blue, the $\left< \Delta \nu \right>$ estimate in red and when implemented, the KBCD approach is in green. The target function in all panels is plotted in black.}
		\label{figmetal}
\end{figure*}  
\begin{table*}[t]
\caption{Inversion results for the test case with a different metallicity, $\mathrm{Model}_{1}$.}
		\label{tabresultsMetal}
  \centering
\begin{tabular}{c | c | c | c}
\hline \hline
 Method & $\bar{\rho}$ $\mathrm{(g/cm^{3})}$& $\tau$ $\mathrm{(s)}$ & $t$ $\mathrm{(s^ {-1})}$\\ \hline 
Reference Value & $2.036$ & $3007.77$ & $-0.002523$\\
SOLA & $2.055$ $\pm 1.17 \times 10^{-4}$ &$2993.91$ $\pm 0.08$&$-0.002548$ $\pm 1.27 \times 10^{-4}$\\ 
$\left< \Delta \nu \right>$ or $\tilde{\delta} \nu$ estimates & $2.054$ $\pm 1.33 \times 10^{-3}$ & $2995.10$ $\pm 0.334$ & $-0.002560$ $\pm 2.71 \times 10^{-5}$\\ 
KBCD & $2.055$ $\pm 4.2 \times 10^{-4}$ & $-$ & $-$\\  
Target Value & $2.047$& $2995.01$ & $-0.002539$\\  \hline
\end{tabular}
\end{table*}
\begin{table*}[t]
\caption{Inversion results for the test case with a different $\alpha_{\mathrm{MLT}}$, $\mathrm{Model}_{2}$.}
		\label{tabresultsAlpha}
\centering
\begin{tabular}{c | c | c | c}
\hline \hline
 Method & $\bar{\rho}$ $\mathrm{(g/cm^{3})}$& $\tau$ $\mathrm{(s)}$ & $t$ $\mathrm{(s^ {-1})}$\\ \hline
Reference Value & $1.523$ & $3471.91$ & $-0.002452$\\
SOLA & $1.533$ $\pm 9.89 \times 10^{-5}$ &$3460.29$ $\pm 0.1$&$-0.002460$ $\pm 1.38 \times 10^{-4}$\\ 
$\left< \Delta \nu \right>$ or $\tilde{\delta} \nu$ estimates & $1.530$ $\pm 9.95 \times 10^{-4}$ & $3464.43$ $\pm 0.45$ & $-0.002464$ $\pm 2.835 \times 10^{-5}$\\ 
KBCD & $1.534$ $\pm 3.14 \times 10^{-4}$ & $-$ & $-$\\  
Target Value & $1.533$& $3461.49$ & $-0.002458$\\  \hline
\end{tabular}
\end{table*}
First of all, we can tell that the inversion of the age indicator is far more accurate when there are no metallicity effects. Indeed, modifying the metallicity affects the entire star, whereas changing the mixing-length only influences the convective envelope, thereby having a negligible impact on the age indicator inversions. Furthermore, test cases carried out for this model with up to $50$ or $70$ frequencies showed an improvement in the accuracy of the method. 
The inversion step, as well as the estimate based on the large frequency separation should only be considered if there's a sufficient number frequencies with small error bars. If this is not the case, then one should avoid carrying out an inversion. We will discuss more extensively the observed weaknesses of the method and possible problems in Sect. \ref{conclusion}. Two supplementary results can be observed for this first test case: the SOLA method is again more accurate when dealing with surface effects, here the variations of $\alpha_{\mathrm{MLT}}$, confirming what had been guessed from the results of the previous section. The second comment is related to the estimates based on the frequency separations. We see that the results improve even if we already fitted the arithmetic average of the large separation during the forward modelling process. This means that the $\chi^{2}$ large separation is more efficient at obtaining the acoustic radius and the mean density of a star and should be preferred over the average large separation. The case of the age indicator is also different since the estimate is determined through the combination given in Eq. (\ref{eqfreqcomb}) and not the small separation alone.
\subsection{Test case with non-adiabatic frequencies}
In this section we present the results for a $0.9$ $M_{\odot}$ model, denoted target $1_{\mathrm{nad}}$, for which non-adiabatic effects have been taken into account. The frequencies have been calculated with the MAD oscillation code, using a non-local, time-dependent treatment of convection taking into account the variations of the convective flux and of the turbulent pressure due to the oscillations \citep[see][for the description of this treatment]{Grigahcene,Dupret, Dupretconv}. A second test case was carried out using a $1$ $M_{\odot}$ model, denoted target $2_{\mathrm{nad}}$, and a slightly less accurate fitting model. The characteristics of both targets are summarised in Table \ref{tabcaractargNadiab}. In both test cases, the difference between the frequencies from the target and reference models lay in the fact that only the former includes non-adiabatic effects. The results are summarised in Tables \ref{tabresultsNadiab2} and \ref{tabresultsNadiab1} for both targets. The kernels from the various inversions and estimates are illustrated in Fig. \ref{figNad2}. The accuracy of the results is clearly related to how well the kernels match their target functions, thereby accounting for the reliability of the inversion technique. We observe that the SOLA inversion technique leads to accurate results for all characteristics in the first test case. For the second test case, we first carried out inversions and estimates based on a set of $33$ frequencies. The results were accurate for the mean density and the acoustic radius. However, the age indicator estimate was as accurate as the value obtained through the forward modelling because the inversion over-corrected this value. Therefore, we carried out a second set of inversions, using $40$ frequencies ranging from $n=15-28$ for $\ell=0$ and from $n=15-27$ for $\ell=1,2$ to see if the result for the age indicator could be improved. This second test is presented in Table \ref{tabresultsNadiab1} where we can see that the SOLA inversion leads to more accurate results than all of the other techniques. This illustrates two effects: firstly, when the model and the target are less well fitted, the inversion requires more frequencies to reach a good accuracy; secondly, a few more frequencies can improve a lot the accuracy of the inversion. This second effect is typical of ill-posed problems. One has to be aware that the accurate result for the second frequency set does not mean that using $40$ frequencies is sufficient in all cases. Analysing the different contributions to the error showed that in this case, the estimates based on frequency combinations could not accurately reproduce non-adiabatic effects in the frequencies. We can thus conclude that the SOLA method is optimal to correct the errors introduced in the forward modelling and particulary surface effects. 
\begin{table}[t]
\caption{Characteristics of targets $1_{\mathrm{nad}}$ and $2_{\mathrm{nad}}$.}
		\label{tabcaractargNadiab}
  \centering
\begin{tabular}{r | c | c }
\hline \hline
 & \textbf{Model $1_{\mathrm{nad}}$}& \textbf{Model $2_{\mathrm{nad}}$}\\ \hline
Mass ($\mathrm{M_{\odot}}$)& $0.9$ & $1.0$  \\
Radius ($\mathrm{R_{\odot}}$) & $0.858$ & $0.942$ \\ 
Age ($\mathrm{Gyr}$) &$6.0$ &$3.0$  \\ 
$T_{\mathrm{eff}}$ ($\mathrm{K}$) & $5335$ & $5649$ \\ 
$\log(g)$ ($\mathrm{dex}$) & $4.5248$ & $4.4895$ \\
$Z$ & $0.0135$ & $0.0135$ \\
$\alpha_{\mathrm{MLT}}$ & $1.62$ & $1.62$ \\ \hline
\end{tabular}
\end{table}
\begin{table*}[t]
\caption{Inversion results for test case $1$, $\mathrm{Model}_{1\mathrm{nad}}$, using $33$ non-adiabatic frequencies.}
\label{tabresultsNadiab2}
  \centering
\begin{tabular}{c | c | c | c}
\hline \hline
 Method & $\bar{\rho}$ $\mathrm{(g/cm^{3})}$& $\tau$ $\mathrm{(s)}$ & $t$ $\mathrm{(s^ {-1})}$\\ \hline
Reference Value & $1.986$ & $3042.76$ & $-0.001873$\\
SOLA & $2.01$ $\pm 1.15 \times 10^{-4}$ &$3024.60$ $\pm 0.08$&$-0.001893$ $\pm 7.8 \times 10^{-5}$\\ 
$\left<\Delta \nu \right>$ or $\tilde{\delta} \nu$ estimates & $1.986$ $\pm 1.3 \times 10^{-3}$ & $3042.80$ $\pm 0.34$ & $-0.001903$ $\pm 2.56 \times 10^{-5}$\\ 
KBCD & $2.015$ $\pm 4.1 \times 10^{-4}$ & $-$ & $-$\\  
Target Value & $2.006$& $3023.88$ & $-0.001894$\\  \hline
\end{tabular}
\end{table*}
\begin{table*}[t]
\caption{Inversion results for test case $2$,$\mathrm{Model}_{2\mathrm{nad}}$ , using $40$ non-adiabatic frequencies.}
\label{tabresultsNadiab1}
  \centering
\begin{tabular}{c | c | c | c}
\hline \hline
 Method & $\bar{\rho}$ $\mathrm{(g/cm^{3})}$& $\tau$ $\mathrm{(s)}$ & $t$ $\mathrm{(s^ {-1})}$\\ \hline
Reference Value & $1.588$ & $3399.79$ & $-0.002285$\\
SOLA & $1.691$ $\pm 9.4 \times 10^{-5}$ &$3294.84$ $\pm 0.09$&$-0.002150$ $\pm 8.5 \times 10^{-5}$\\ 
$\left< \Delta \nu \right>$ or $\tilde{\delta} \nu$ estimates & $1.659$ $\pm 7.9 \times 10^{-4}$ & $3326.93$ $\pm0.3$ & $-0.002248$ $\pm 2.54 \times 10^{-5}$\\ 
KBCD & $1.696$ $\pm 2.65 \times 10^{-4}$ & $-$ & $-$\\  
Target Value & $1.684$& $3295.87$ & $-0.002190$\\  \hline
\end{tabular}
\end{table*}
\subsection{Test case with turbulent pressure}
In the last test case, we included the effects of turbulent pressure when calculating, thanks to the LOSC code, the adiabatic pulsation frequencies of a $1$ $\mathrm{M}_{\odot}$ target. The turbulent pressure was included in the computation of the evolution of the model by adding a supplementary term $P_{\mathrm{turb}}$ using the following phenomenological approach:
\begin{align}
P_{\mathrm{turb}}=\left< \rho v_{R}^{2} \right> = C_{p_{\mathrm{turb}}} \rho v_{R}^{2},
\end{align} 
with $v_{R}$ the radial speed of the convective elements given by the mixing length theory. The value of the turbulent pressure coefficient $C_{p_{\mathrm{turb}}}$ was chosen to be $1.58$ to match effects of $3\mathrm{D}$ simulations for the sun. The characteristics of the target are summarised in Table \ref{tabcaractargtestturb} and the results are summarised in Table \ref{tabresultsturb}. We see that the SOLA method can account for the effects of turbulent pressure and improve the accuracy with which global stellar characteristics are determined in this case. The kernels for this inversion are illustrated in Fig. \ref{figsturb}. As was the case previously, the SOLA kernels seem to be more regular and closer to their target functions than those of the other techniques. 
\begin{table}[t]
\caption{Characteristics of target $1_{\mathrm{turb}}$.}
\label{tabcaractargtestturb}
  \centering
\begin{tabular}{r | c}
\hline
 & \textbf{Model $1_{\mathrm{turb}}$}\\ \hline
Mass ($\mathrm{M_{\odot}}$)& $1.0$  \\
Radius ($\mathrm{R_{\odot}}$) & $0.868$ \\ 
Age ($\mathrm{Gyr}$) &$4.0$  \\ 
$T_{\mathrm{eff}}$ ($\mathrm{K}$) & $5683$  \\ 
$\log(g)$ ($\mathrm{dex}$) & $4.469$  \\
$Z$ & $0.0135$  \\
$\alpha_{\mathrm{MLT}}$ & $1.62$  \\ \hline
\end{tabular}
\end{table}
\begin{table*}[t]
\caption{Inversion results for test case using turbulence pressure, $\mathrm{Model}_{1\mathrm{turb}}$.}
\label{tabresultsturb}
  \centering
\begin{tabular}{c | c | c | c}
\hline \hline
 Method & $\bar{\rho}$ $(\mathrm{g/cm^{3}})$& $\tau$ $\mathrm{(s)}$ & $t$ $\mathrm{(s^ {-1})}$\\ \hline
Reference Value & $1.557$ & $3429.59$ & $-0.001877$\\
SOLA & $1.575$ $\pm 9.6 \times 10^{-5}$ &$3409.72$ $\pm 0.1$&$-0.001894$ $\pm 6.7 \times 10^{-5}$\\ 
$\left< \Delta \nu \right>$ or $\tilde{\delta} \nu$ estimates & $1.570$ $\pm 1.02 \times 10^{-3}$ & $3415.93$ $\pm0.4$ & $-0.001902$ $\pm 2.6 \times 10^{-5}$\\ 
KBCD & $1.576$ $3.3 \times 10^{-4}$ & $-$ & $-$\\  
Target Value & $1.573$& $3409.76$ & $-0.001888$\\  \hline
\end{tabular}
\end{table*}
\section{Conclusion} \label{conclusion}
In this article, we have analysed four different methods for obtaining various stellar parameters. These include: asymptotic relations based on two different implementations of the large and small frequency separations, a scaling law for the mean density which includes the \citet{Kjeldsen} surface corrections, and inversions based on the SOLA method. A comparison of these different
methods reveals the following strengths and weaknesses:
\begin{itemize}
\item \textbf{Arithmetic average of the large and small frequency separations}: this method is the simplest to implement and is useful in the forward modelling. It is, however, less accurate than the other methods.
\item \textbf{Large frequency separation from a $\chi^2$ adjustment, and arithmetic average of the age indicator (based on Eq.
 \ref{eqfreqcomb})}: this remains simple but is more accurate than the previous
approach, as demonstrated by the improvement in the results when this method is applied after the forward modelling (which uses the previous approach). The reason why this version of the large frequency separation is more accurate is because it uses the information from all of the modes, rather than simply the ones with the lowest and highest n values. The reason why using the average age indicator works better than the average small frequency separation is less obvious but is likely to be related to the fact that in the former case one isolates an integral which only depends on the stellar structure and does not contain a mode-dependant coefficient in front, before carrying out the
average. In spite of these improvements, this approach remains sensitive to surface effects as shown, for instance, in Fig. \ref{figresultau} (left column).
\item \textbf{The mean density from the Kjeldsen et al. (2008) surface-correcting approach}: This approach produces superior results compared to the two previous methods because it is able to correct for surface effects. However, changes in metallicity affect both this method and SOLA inversions more than the previous methods, since such changes modify the entire star rather than just the near-surface layers.
\item \textbf{SOLA inversions}: although this approach is the most complicated, it also turns out to be the most accurate. Indeed, apart from the case where the metallicity was modified, it is able to deal with incorrect assumptions in the reference models since it focusses on optimising the averaging and cross-term kernels. Furthermore, a key feature of SOLA inversions is that the quality
of these kernels is closely related to the quality of the results, unlike what sometimes happens for scaling laws where fortuitous compensations lead to good results. This is important because it gives a way of estimating the quality of the inversion results. However, we do note that one must be careful to choose a reference model which is sufficiently close to the target, particularly for the age indicator inversions. This naturally leads to the use of forward modelling before application of this method. A quick inspection of the values in Tables \ref{tabresultsAlpha},\ref{tabresultsMetal},\ref{tabresultsNadiab1},\ref{tabresultsNadiab2},\ref{tabresultsturb} shows that SOLA inversions have improved the accuracy by a factor ranging from $10$ to several hundred for $\tau$ and $\bar{\rho}$ and from $1.125$ to more than $20$ for $t$, when compared to results from the forward modelling.
\\
\\
A couple of further comments need to be made concerning SOLA inversions of the age indicator. Firstly, great care should be taken when calculating the quantity $\frac{1}{x}(\frac{dc}{dx})$, which intervenes in the
target function. Indeed, this quantity is prone to numerical noise as $x$ approaches $0$. In our calculations, we reduced such noise by numerically calculating the derivative with respect to $x^{2}$, but note that it was still necessary to inspect this function before carrying out the inversion. Secondly,
as can be seen from the top left panel of Fig. \ref{figsmallsepkernels}, the target function does not go to $0$ in the centre, as opposed to the structural kernels which behave as $\mathcal{O}(r^{2})$ in the centre. Therefore this target will be difficult to fit, even with more frequencies, and we need to find a workaround to be able to retrieve the effects of stellar evolution with an inversion technique. In fact, the lower left panel of Fig. \ref{figsmallsepkernels} shows that optimising the anti-derivative is not always sufficient to solve this problem. 
\\
\\
In future studies, we plan to analyse in more detail under what conditions SOLA inversions yield good results. In particular, we will investigate, in a systematic way, how close the reference model needs to be to the observed star for the inversion to be reliable. It will also be important to test the quality of the averaging and cross-term kernels as a function of the number and type of modes available. We also plan to extend SOLA inversions to other structural quantities, including age indicators which do not suffer from the difficulties mentioned above. This highlights the great potential of the SOLA method, since it allows us to choose the global structural characteristic that we wish to determine, offering a promising new diagnostic method into stellar structural properties.
\end{itemize}
\begin{acknowledgements}
We thank the referee for helpful comments and suggestions
which have improved the manuscript. G.B. is supported by the FNRS (``Fonds National de la Recherche Scientifique'') through a FRIA (``Fonds pour la Formation à la Recherche dans l'Industrie et l'Agriculture'') doctoral fellowship. D.R.R. is currently funded by the European Community's Seventh Framework Programme (FP7/2007-2013) under grant
agreement no. 312844 (SPACEINN), which is gratefully acknowledged. 
\end{acknowledgements}
\appendix
\section{Demonstration of the error propagation formula for the non-linear extension of the acoustic radius inversion}\label{secapperror}
Equation (\ref{errornonlinear}) is obtained with a little algebra. First, we treat the observed frequencies, $\nu_{obs,i}$, and the inverted acoustic radius, $\tau_{inv}$, as independent stochastic variables:
\begin{align}
\nu_{\mathrm{obs},i}=\bar{\nu}_{\mathrm{obs},i}(1+\epsilon_{i}), \\
\tau_{\mathrm{inv}}= \bar{\tau}_{\mathrm{inv}}(1+\epsilon_{\tau}),
\end{align}
with $\epsilon_{i}$ being the individual noise realisations for each frequency, $\epsilon_{\tau}$ the resultant deviation on $\tau_{\mathrm{inv}}$, and $\bar{\nu}_{\mathrm{obs}}$ and $\bar{\tau}_{\mathrm{inv}}$ the average of the stochastic variables $\nu_{\mathrm{obs},i}$ and $\tau_{\mathrm{inv}}$, respectively.. Furthermore, we assume that: 
\begin{align}
\epsilon_{i} \ll 1.
\end{align}
Using the fact that $\tau_{\mathrm{inv}}=q_{\mathrm{opt}} \tau_{\mathrm{ref}}$ with the definition of $q_{opt}$ given in Eq. (\ref{optimaltau}) and the separation into stochastic and average contributions defined previously, we get:
\begin{align}
\bar{\tau}_{\mathrm{inv}}(1+\epsilon_{\tau})& \simeq \frac{-\tau_{\mathrm{ref}}}{\sum_{i}c_{i}\dfrac{\bar{\nu}_{\mathrm{obs},i}}{\nu_{\mathrm{ref},i}}\left[1+\frac{\sum_{i}c_{i}\dfrac{\bar{\nu}_{\mathrm{obs},i}}{\nu_{\mathrm{ref},i}}\epsilon_{i}}{\sum_{i}c_{i}\dfrac{\bar{\nu}_{\mathrm{obs},i}}{\nu_{\mathrm{ref},i}}}\right]}\nonumber \\
&=\frac{-\tau_{\mathrm{ref}}}{\sum_{i}c_{i}\dfrac{\bar{\nu}_{\mathrm{obs},i}}{\nu_{\mathrm{ref},i}}}\left[1- \frac{\sum_{i}c_{i}\dfrac{\bar{\nu}_{\mathrm{obs},i}}{\nu_{\mathrm{ref},i}}\epsilon_{i}}{\sum_{i}c_{i}\dfrac{\bar{\nu}_{\mathrm{obs},i}}{\nu_{\mathrm{ref},i}}} \right],
\end{align}
where we assumed that $\epsilon_{i}$ is much smaller than 1, thereby allowing us to linearise the above equation. We now apply the formula for the variance of a linear combination of independant stochastic variables and obtain:
\begin{align}
\bar{\tau}_{\mathrm{inv}}^{2}\sigma_{\tau}^{2}=\frac{\tau_{\mathrm{ref}}^{2}}{\left(\sum_{i}c_{i}\dfrac{\bar{\nu}_{\mathrm{obs},i}}{\nu_{\mathrm{ref},i}}\right)^{4}}\sum_{i}c_{i}^{2}\sigma_{i}^{2},
\label{eqoptimaltauerror}
\end{align}
where we used the following equivalences:
\begin{align}
\sigma^{2}_{i}&=\sigma^{2}_{\frac{\delta \nu_{i}}{\nu_{i}}}=\sigma^{2}_{\frac{\nu_{\mathrm{obs},i}}{\nu_{\mathrm{ref},i}}} \nonumber \\
&= \left(\frac{\bar{\nu}_{\mathrm{obs},i}}{\nu_{\mathrm{ref},i}}\right)^{2}\sigma^{2}_{\epsilon_{i}}
\end{align}  
Equation \ref{eqoptimaltauerror} then leads directly to Eq. (\ref{errornonlinear}) when using the definition of $q_{\mathrm{opt}}$ given in Eq. (\ref{optimaltau}). 
\section{Supplementary figures}
The following figures illustrate the quality of the kernel fits for some of the test cases we presented in the article. Although these plots are redundant on the visual point of view, we wish here again to stress that they are crucial to the understanding of the quality of a SOLA inversion and justify the accuracy of the results presented in the previous sections. 
\begin{figure*}[!h]
  \centering
		\includegraphics[width=15cm]{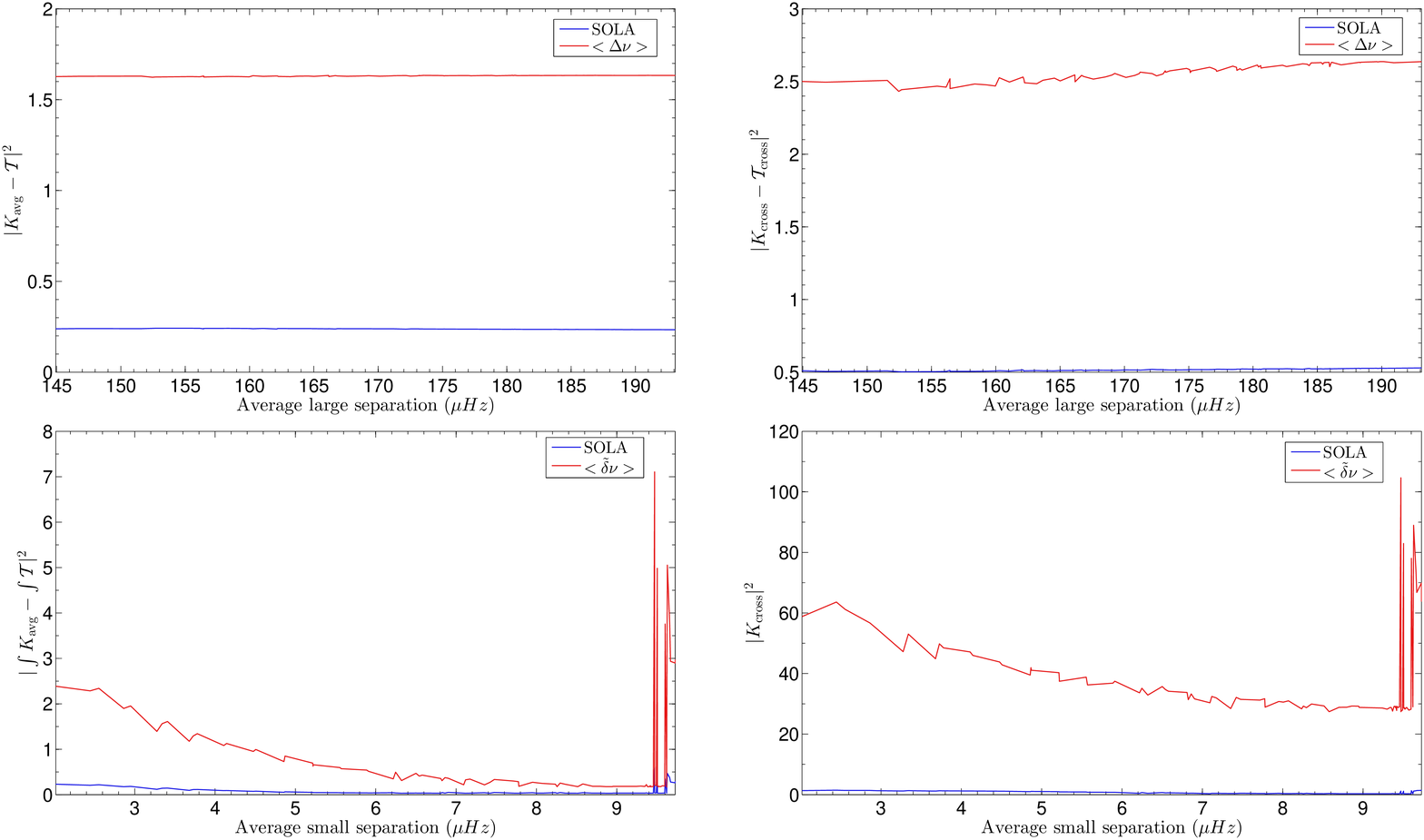}
	\caption{Least square fits of the kernels for model $A^{'}$.}
	\vspace{0.3cm}
		\label{figKerFits}
\end{figure*} 

\begin{figure*}[!h]
  \centering
    \textbf{Kernels for $\tau$, $\bar{\rho}$ and $t$ for Model $1_{\mathrm{nad}}$}
		\includegraphics[width=15cm]{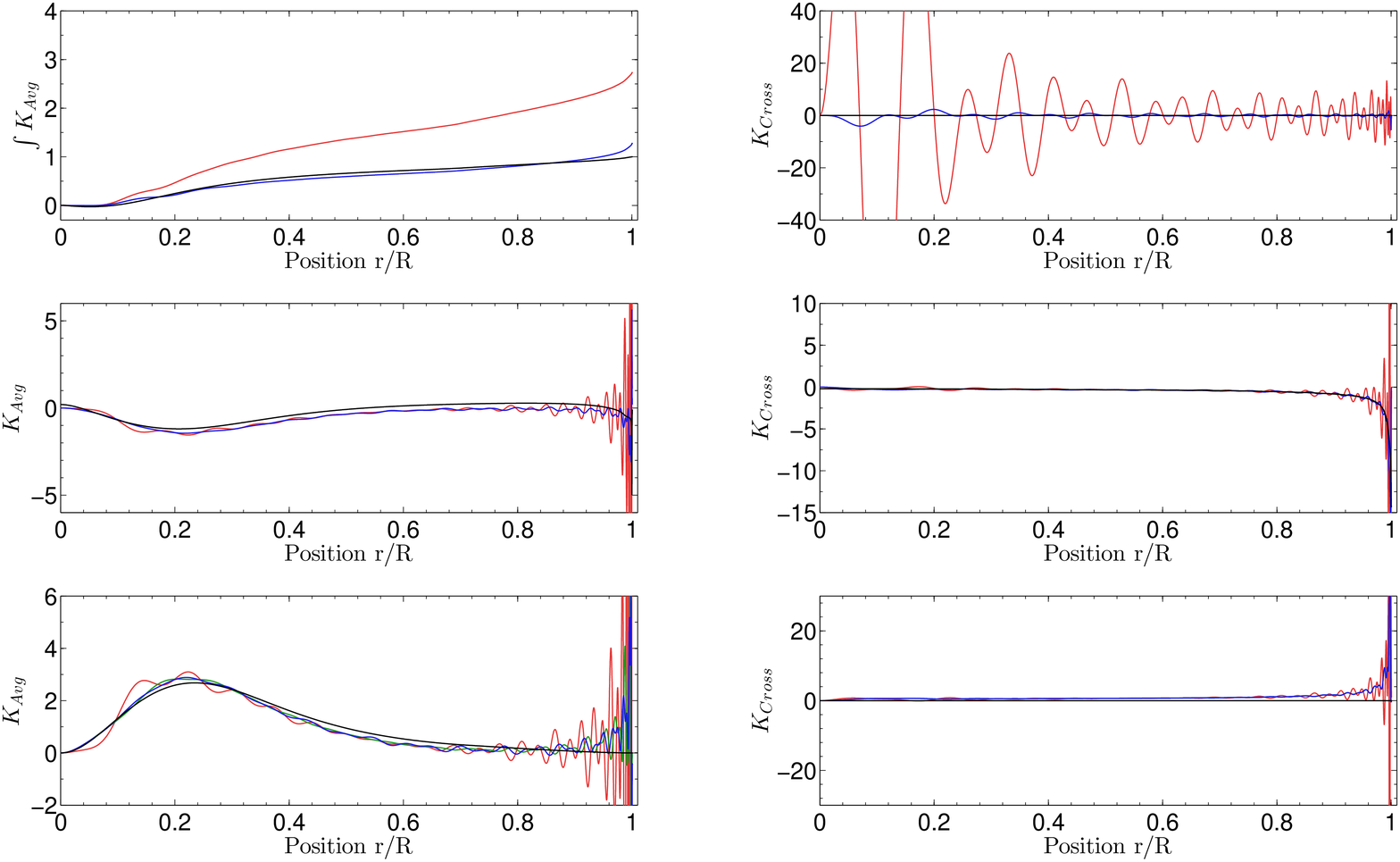}
	\caption{(Colour online) same as Fig. \ref{figmetal} for the first test case with non-adiabatic frequencies.}
	\vspace{0.3cm}
		\label{figNad2}
    \textbf{Kernels for $\tau$, $\bar{\rho}$ and $t$ for Model $1_{\mathrm{turb}}$}
		\includegraphics[width=15cm]{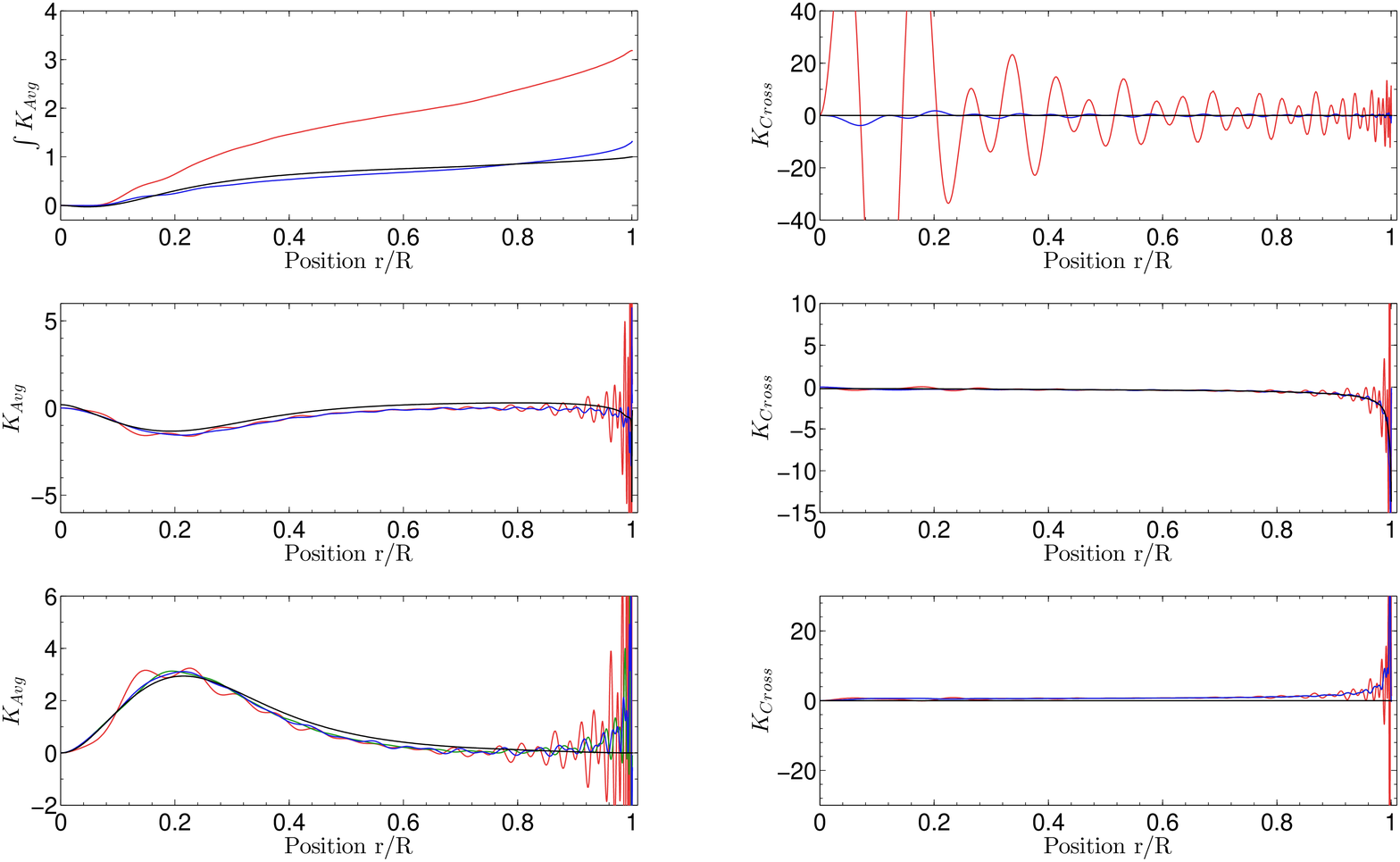}
	\caption{(Colour online) Same as Fig. \ref{figmetal} for the test case with turbulent pressure.}
		\label{figsturb}
\end{figure*} 

\bibliography{biblioarticle}
\end{document}